\documentclass[onecolumn, numberedappendix]{aastex6}
\pdfoutput=1
\usepackage{graphicx}
\usepackage{epsfig}
\usepackage{natbib}
\usepackage{amssymb}
\usepackage{amsbsy}
\usepackage{natbib}
\usepackage{url}
\usepackage[mathcal]{euscript}
\bibpunct{(}{)}{,}{a}{}{,}

\newcommand{\msun}{M$_{\sun}$}

\newcommand{\etal}{et~al.}

\newcommand{\vmk}{$(V-K_{\rm s})_0$}

\begin{document}
 
\def\simlt{\vcenter{\hbox{$<$}\offinterlineskip\hbox{$\sim$}}}
\def\simgt{\vcenter{\hbox{$>$}\offinterlineskip\hbox{$\sim$}}}
\def\kms{km s$^{-1}$}

\title{The Rotational Evolution of Young, Binary M Dwarfs}
\author{John Stauffer\altaffilmark{1},
Luisa M. Rebull\altaffilmark{2,1},
Ann Marie Cody\altaffilmark{3},
Lynne A. Hillenbrand\altaffilmark{4},
Marc Pinsonneault\altaffilmark{5},
David Barrado\altaffilmark{6},
Jerome Bouvier\altaffilmark{7},
Trevor David \altaffilmark{8}
}
\altaffiltext{1}{Spitzer Science Center (SSC), IPAC, California Institute of
Technology, Pasadena, CA 91125, USA; stauffer@ipac.caltech.edu}
\altaffiltext{2}{Infrared Science Archive (IRSA), IPAC, California
Institute of Technology, 1200 E. California Blvd,
MS 100-22, Pasadena, CA 91125 USA}
\altaffiltext{3}{NASA Ames Research Center, Space Sciences and
Astrobiology Division, MS245-3, Moffett Field, CA 94035 USA}
\altaffiltext{4}{Astronomy Department,
California Institute of Technology, Pasadena, CA 91125 USA}
\altaffiltext{5}{Astronomy Department, The Ohio State University,
Columbus, OH  43210}
\altaffiltext{6}{Centro de Astrobiolog\'ia, Dpto. de
Astrof\'isica, INTA-CSIC, E-28692, ESAC Campus, Villanueva de
la Ca\~nada, Madrid, Spain}
\altaffiltext{7}{Univ. Grenoble Alpes, IPAG, 38000, Grenoble, France}
\altaffiltext{8}{Jet Propulsion Laboratory, California Institute of Technology, 
M/S 321-100, 4800 Oak Grove Drive, Pasadena, CA 91109, USA}

\begin{abstract}

We have analysed K2 light curves for more than 3,000 low mass stars in
the $\sim$8 Myr old Upper Sco association, the $\sim$125 Myr age
Pleiades open cluster and the $\sim$700 Myr old Hyades and Praesepe
open clusters to determine stellar rotation rates.   Many of these K2
targets show two distinct periods, and for the lowest mass stars in
these clusters virtually all of these systems with two periods are
photometric binaries. The most likely explanation is that we  are
detecting the rotation periods for both components of these binaries. 
We explore the evolution of the rotation rate in both components of
photometric binaries relative to one another and to  non-photometric
binary stars. In Upper Sco and the Pleiades, these low mass binary
stars have periods that are much shorter on average and much closer to
each other than would be true if drawn at random from the M dwarf
single stars.   In Upper Sco, this difference correlates strongly with
the presence or absence of infrared excesses due to primordial
circumstellar disks -- the single star population includes many stars
with disks, and their rotation periods are distinctively longer on
average than their binary star cousins of the same mass.   By Praesepe
age, the significance of the difference in rotation  rate between the
single and binary low mass dMs is much less, suggesting that angular
momentum loss from  winds for fully-convective zero-age main sequence
stars erases memory of the rotation rate dichotomy for binary and
single very low mass stars at later ages.

\end{abstract}

\section{Introduction}

There have been a number of observational programs aimed at
determining the initial distribution of rotation rates for low mass
stars (e.g., Hartmann \etal\ 1986; Bouvier \etal\ 1986; Herbst \etal\
2001; Rebull \etal\ 2002; Moraux \etal\ 2013; Affer \etal\ 2013).
Quite a few observational papers have also been devoted to determining
the multiplicity function for low mass stars, both in young
populations (e.g. Ghez \etal\ 1993; Bouvier \etal\ 1997; Kraus \etal\
2008;  Shan \etal\ 2017) and amongst field stars of indeterminate age
(e.g., Duquennoy \& Mayor 1991; Fischer \& Marcy 1992; Reid \& Gizis
1997; Janson \etal\ 2012).  However, it has seldom if ever been
possible to conduct sensitive surveys with large sample sizes which
simultaneously place strong constraints on both rotation rates and
multiplicity.   If such a survey could be conducted, it could
potentially provide valuable new insights into how binary (and single)
stars are formed and evolve.

In fact, a facility which could simultaneously identify binary stars
and measure rotation rates for large populations of young low mass
stars in nearby open clusters and star-forming regions did become
available in 2014. That facility is NASA's K2 space telescope (Howell
\etal\ 2014).   K2 has now provided high precision, rapid cadence, 
long-duration, sensitive light curves for a large fraction of the
members in several of the nearest and best studied nearby star-forming
regions and open clusters. Those light curves not only allow
determination of the rotation periods for a very large percentage of
the target stars, but they also provide rotation periods for both
components of many of the binary systems (i.e. where there is enough light 
from the secondary to detect its rotational modulation as well as that 
of the primary star).  Because the ages for these
clusters are well determined, it should be possible to use these data
to place new empirical constraints on the evolution of rotation and
multiplicity with time.

In this paper, we use our K2 rotational periods for the Upper Sco
association (age $\sim$8 Myr), the Pleiades (age $\sim$125 Myr) and
the older (age $\sim$700 Myr) Hyades and Praesepe open clusters  to
highlight possible evidence that the rotation rates and multiplicity
of low mass stars are strongly linked at 8 Myr, with stars in
photometric binary systems being both more rapidly rotating and with
the two components of the binaries having rotation rates closer to
each other than would be true if drawn randomly from the single star
population. In \S 2, we describe the sources of data for this paper.
In \S 3, we document the ability of K2 to determine the rotation
periods for both component stars of binary dM stars. In \S 4, we
compare the rotation rates of the single and binary $M <$ 0.32 \msun\
M dwarfs in Upper Sco, Pleiades, Hyades, and Praesepe, and provide
evidence that the binary and single stars have different rotational
velocity distributions at young ages. In \S 5, we describe a Monte
Carlo simulation which supports our conclusion that the binary dMs in
Upper Sco and the Pleiades are faster rotating and have more closely
paired rotation rates than would be true if drawn from the single dM
population. Finally, in \S 6, we discuss links of these differences to
the presence of pre-main-sequence (PMS) disks and in \S 7 we discuss
some of the properties of the binary populations in Upper Sco and the
Pleiades and how those properties help us interpret the rotation
period data.

\section{Observational Data}

All of the rotation period data we use in this paper come from light
curves obtained by NASA's K2 mission.  We use the periods we have
previously  derived and reported in Rebull \etal\ (2016ab) for the
Pleiades, Rebull \etal\ (2017) for Praesepe, Rebull \etal\ (2018a) for
Upper Sco, and to be reported in an upcoming paper for the Hyades
(Rebull \etal\ in preparation).  In those same papers, we provide
\vmk\ colors for all of the stars with periods.  Where
possible, we use observed $V$ and $K_{\rm s}$ photometry to yield the
\vmk\ color; where that was not possible (primarily when
there is no literature $V$ magnitude; $K_{\rm s}$ comes from 2MASS),
we use photometry at other bands to allow us to estimate the star's
\vmk\ color. Additionally, there is effectively no reddening
towards Pleiades, Hyades, or Praesepe; reddening matters for Upper
Sco, and our method of dereddening is discussed in  Rebull \etal\
(2018a).  The set of stars observed by K2 which we consider to be
members of the four clusters is also described in those papers.  In
general, the membership status is quite good in the Pleiades, Hyades,
and Praesepe, and we expect there are relatively few non-members in
our K2 cluster catalogs.   The situation is considerably worse in
Upper Sco, where perhaps 5-10\% of the 1133 stars in our catalog may
be non-members (Rebull \etal\ 2018a), in addition to the scatter in
color introduced by reddening.

Because we believe the Hyades and Praesepe have very similar
ages, and because we have comparatively few very low mass stars with K2
periods in these two clusters, we combine the period data we have for
them in all of the plots for this paper.  

Upper Sco is young enough that some of its members retain their
primordial circumstellar disks and in many cases those stars are still
actively accreting gas.   Rebull \etal\ (2018a) identified Upper Sco
members with primordial disks by analysing the spectral energy
distributions (SEDs) of those stars; infrared (IR) excesses were
identified primarily using 2MASS (Skrutskie \etal\ 2006), WISE (Wright
\etal\ 2010), and Spitzer (Werner \etal\ 2004) data.  We use the
sorting of Upper Sco stars into disk-bearing and non-disk samples in
\S 6 of this paper.

\section{Setting the Stage - Defining our M Dwarf Binary Star Sample}

Our goal in this paper is to use the K2 rotation data  to determine if
there is a correlation between binarity and rotation for low mass M
dwarfs that originates at birth, and whether that correlation -- if it
exists -- persists through the first Gyr on the main sequence. It is
possible to use knowledge of the observed properties of low mass
binary systems (Duchene \& Kraus 2013) and standard angular momentum
evolution models (e.g., Gallet \& Bouvier 2015) to predict what we
might see.  M dwarf binaries, and particularly low mass M dwarf
binaries are weighted towards relatively small separations (a $<$ 50
AU)  and high-$q$ systems (Bergfors \etal\ 2010; Janson \etal\ 2012)
relative  to their higher mass counterparts. For Upper Sco, high-$q$,
small separation  binaries are likely weighted towards systems without
inner, primordial disks (weak-lined T Tauris, or WTTs) due to the
deleterious effects of  binarity on disk lifetime and mass (Harris
\etal\ 2012; Kraus \etal\ 2016). In the standard disk-locking paradigm
(K\"onigl 1991),  slow rotation is associated with long disk lifetimes;
in the absence of  disks, pre-MS stars should spin up from angular
momentum conservation.  The components of these close binaries should
therefore be relatively rapidly rotating at young ages.  The fact that
the angular momentum loss rate from winds is higher for rapid rotators
than slow rotators should eventually eliminate an overall dichotomy in
rotation rate between binary stars and single stars.  However, because
rapidly rotating dM stars should be in the saturated regime where
angular momentum loss rates are nearly independent of rotation rate,
if both members of a binary are rapid rotators, their rotation rates
could initially diverge from each other. Remarkably,  these
predictions are confirmed by our K2 data.

Before we can conduct this study, we need to establish three things:
\begin{itemize}
\item That our K2 data are capable of identifying a reliable binary
   star population;
\item That there is indeed a built-in preference for rapid rotation 
   amongst binary stars at very young ages;
\item And that we can determine an appropriate mass range over which 
   to compare dM rotation rates for stars ranging in age from $\sim$10 Myr
   to $\sim$1 Gyr.
\end{itemize}

We address each of these points individually in the following sections.

\subsection{Using K2 Light Curves to Identify Low Mass Binary Stars}

\begin{figure}[ht]
\epsscale{0.9}
\plotone{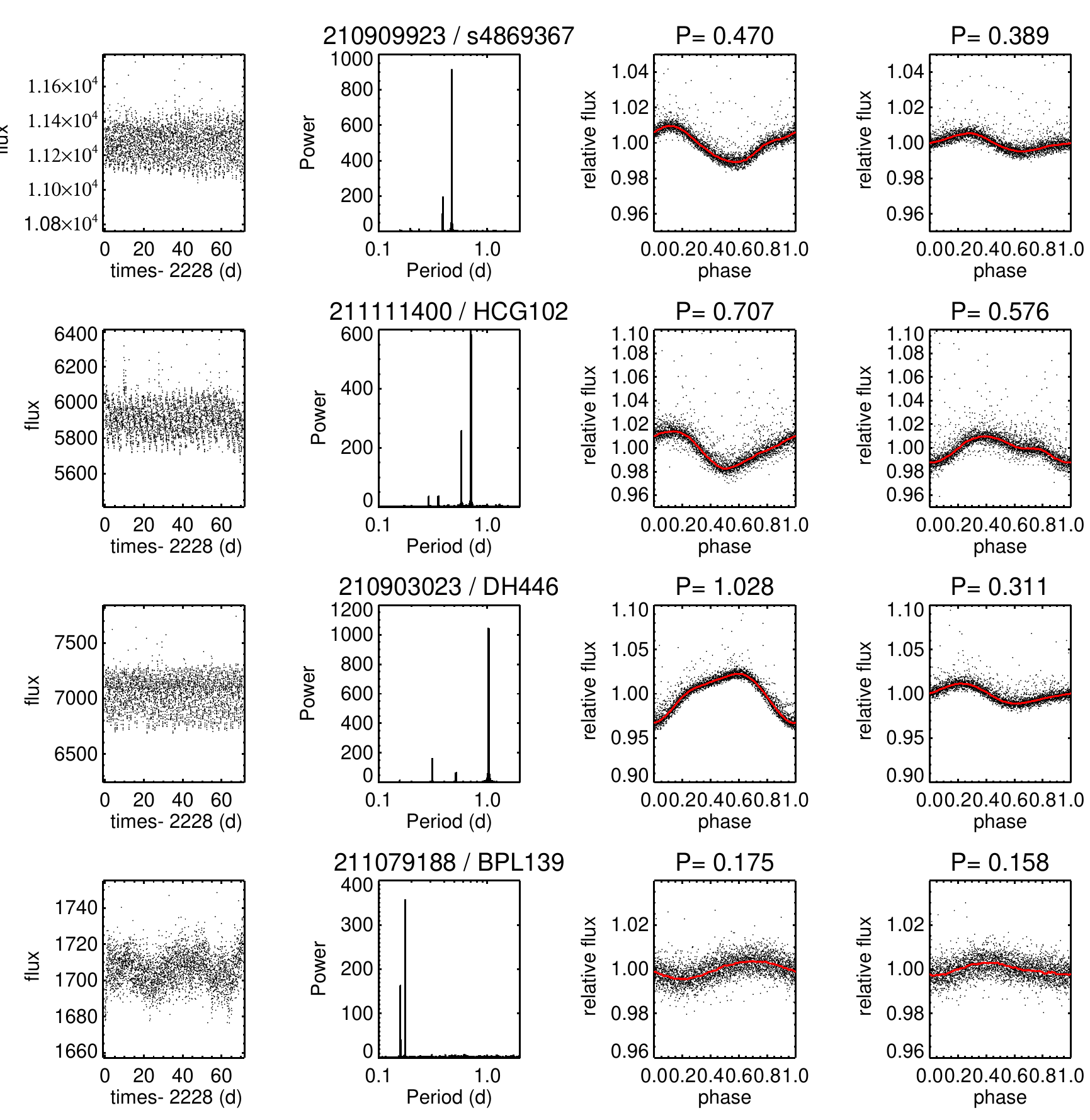} 
\caption{Example light curves (left), LS periodograms (center), and
phased light curves (right) for the individual components of four
Pleiades dM stars (one per line) with two identified rotation periods.
Note that while we can accurately extract the phased light curve shape
and amplitude in counts for each component of these binaries, the
amount to which each component contributes to the total light of the
binary is unknown.  For the phased light curves shown here, we have
simply divided the observed count rate by the median.  
\label{fig:Figure1}}
\end{figure}

In our K2 rotation papers, we have identified photometric periods for
$>$3000 stars in our four clusters (Upper Sco, Pleiades, Hyades, and
Praesepe).  Unlike the case for nearly all ground-based efforts, the
completeness for detecting rotation periods for the low mass stars in
these clusters is very high ($\gtrsim$ 90\%) -- therefore, any
incompleteness correction for pole-on stars or very slowly rotating
stars should have negligible impact on our conclusions\footnote{This
completeness estimate is simply the ratio of the number of candidate
low mass members with a K2 period to the total number of candidate
members in our input K2 cluster sample.  Some of the stars with no
period could be older, field stars mistakenly identified as cluster
members.  The completeness estimate also does not take binary stars
into consideration; that topic is discussed in detail in Appendix
C.}.   For most of our target stars, there is only one real period
identified in the Lomb-Scargle (LS) periodogram.  However, the K2
light curves are so sensitive and have such good cadence and such long
durations that we are often able to detect more than one period for a
given  star when such multiple periods are present.  For stars with
mass $>$1 \msun, multiple periods normally arise from pulsation
(Zwintz \etal\ 2014; White \etal\ 2017).    For 0.5 $<M<$ 1.0 \msun,
multiple periods or complex/structured peaks in the periodogram are
most often due to latitudinal differential rotation and spots at
different latitudes and/or spot evolution (Aigrain \etal\ 2015; Santos
\etal\ 2017; Rebull \etal\ 2016a,b); some stars with two LS peaks in
this mass range are instead binaries, with each period corresponding
to the rotation period of one of the stars in the system.   The latter
option becomes the more likely scenario when the difference between
the two periods is greater than about 20\% (i.e., larger than could be
explained for a plausible degree of latitudinal differential
rotation).   For $M <$ 0.5 \msun, particularly for fully-convective M
dwarfs, both observations and theory suggest that there should be
little or no latitudinal differential rotation (Morin \etal\ 2010;
Reinhold \etal\ 2013;  Rappaport \etal\ 2014; Kuker \& Rudiger 1997;
Browning 2008).    That, combined with the rapid rotation of young dM
stars, suggests that in this mass regime detection of two periods
separated by more than a few percent (Rappaport \etal\ 2014) likely 
signals that the star is a binary\footnote{Stars whose phased light
curves have significantly non-sinusoidal shapes can also have multiple
apparent peaks in their LS periodogram whose periods are harmonics of
the rotation period.  It is normally easy to determine that these
additional peaks are aliases rather than true rotation periods from
close examination of the light curve shape and of the periods
themselves.}. Figure \ref{fig:Figure1} shows the K2 light curves of
four dM K2 binary members of the Pleiades,  their LS periodograms, and
the phased light curves of both component  stars.  Here and throughout
the rest of the paper, we use the short-hand of ``K2 binary" to mean a
low mass star whose K2 light curve shows two peaks in its LS
periodogram, which we interpret to mean that it is a binary star and
that each peak is the rotation period for one component of the
binary.  We use the term ``single" for stars with only one period, but
they could be binaries with low q or tidally-locked  systems.  The
phased light curves of all eight of these stars have shapes that are
typical of what is seen for the single dM members of the cluster.  

\begin{figure}[ht]
\epsscale{0.9}
\plotone{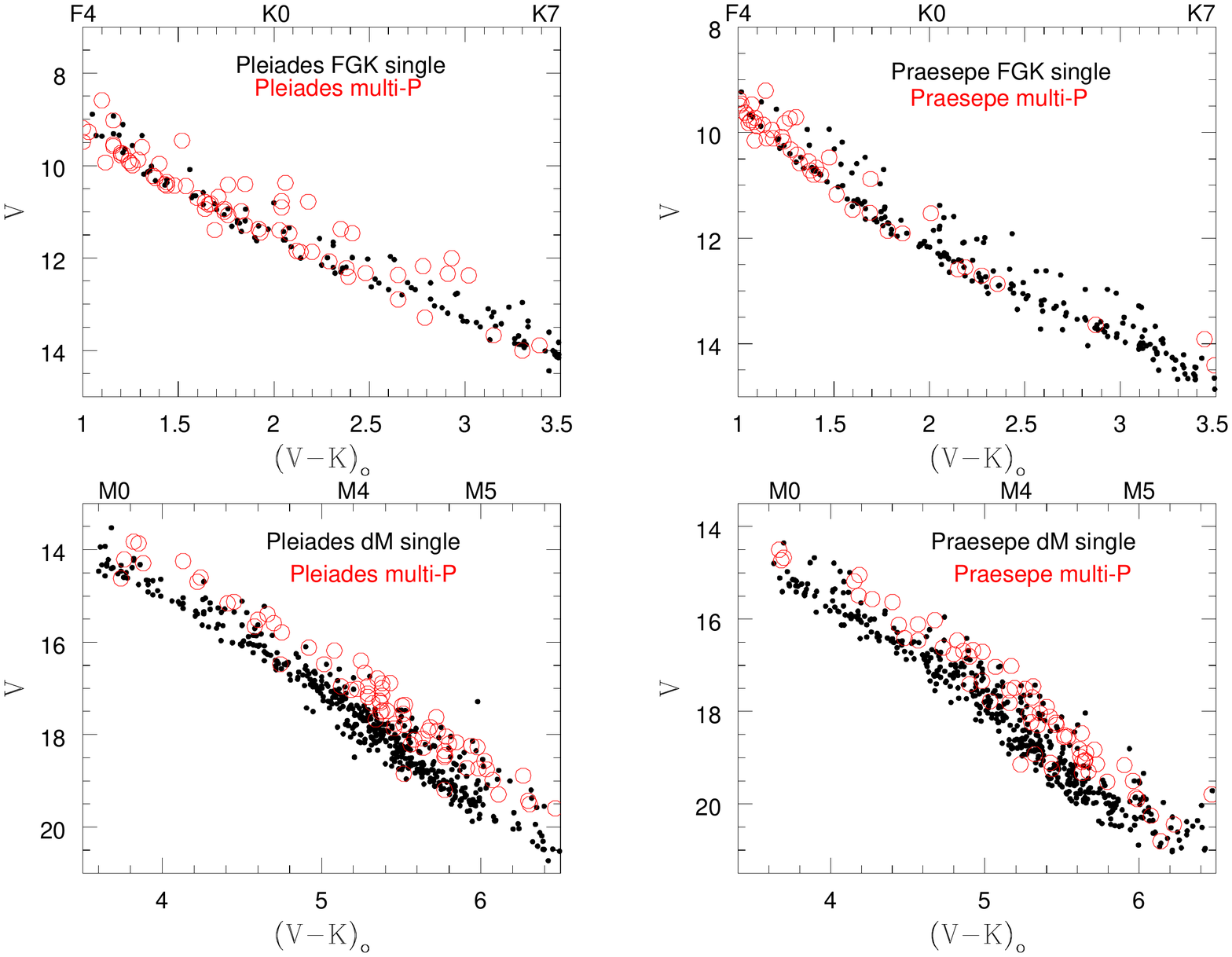}
\caption{CMDs for two mass ranges in the Pleiades (left) and Praesepe
(right).  The top row has stars with $M >$ 0.5 \msun; the bottom row
has stars with $M <$ 0.5 \msun. In all four panels, stars with just
one period in their K2 data are shown as black dots, and stars with
more than one period are shown as open, red circles. For $(V-K_{\rm
s})_0 <$ 3.5, there is little or no preference for the stars with
multiple periods to lie within the expected locus of single or binary
stars.   However, for $(V-K_{\rm s})_0>$ 3.5, the stars with more than
one period are with very few exceptions displaced well above the
single star locus -- thus supporting interpretation that these are
indeed binary stars.
\label{fig:Figure2}}
\end{figure}

Strong confirmation that the dM stars in young clusters with two K2
periods are best interpreted as  high-$q$ binaries is provided in
Figure \ref{fig:Figure2}, where we show color-magnitude diagrams
(CMDs) for our K2 stars for the Pleiades and Praesepe.  The K2 dM
binary stars are nearly always well-displaced  above the single-star
locus in these two diagrams, with a mean $\Delta V >$ 0.5 mag. 
Similar CMDs are not as well-defined for Upper Sco because the
uncertainty in dereddened color is fairly large, but a clear net
displacement between the dM stars with one period and those with two
periods is evident (see Figure 13 of Rebull \etal\ 2018a).

\subsection{Rotation of M Dwarfs in Upper Sco - Rapid Rotation in Binary dMs
    Is Built-in from Early in PMS Evolution}

In Stauffer \etal\ (2016), we compared the periods for mid-dM
($\sim$M4 to $\sim$M6) Pleiades stars identified as binaries by K2 to
the periods for the Pleiades K2 single mid-dM stars.   We found that
the components of the binary dM systems rotated faster on average, and
had periods that were closer to each other, than would have been
expected if those periods were drawn at random from the Pleiades
single dM population.  A Kolmogorov-Smirnov (KS) test and Monte-Carlo
simulations showed that those differences were significant at better
than 1 in 1000 odds.   Other papers reporting evidence that binary,
low mass zero-age main sequence (ZAMS) stars rotate faster than
singles of the same mass include Meibom \etal\ (2007) and Douglas
\etal\ (2016, 2017).

\begin{figure}[ht]
\epsscale{0.9}
\plotone{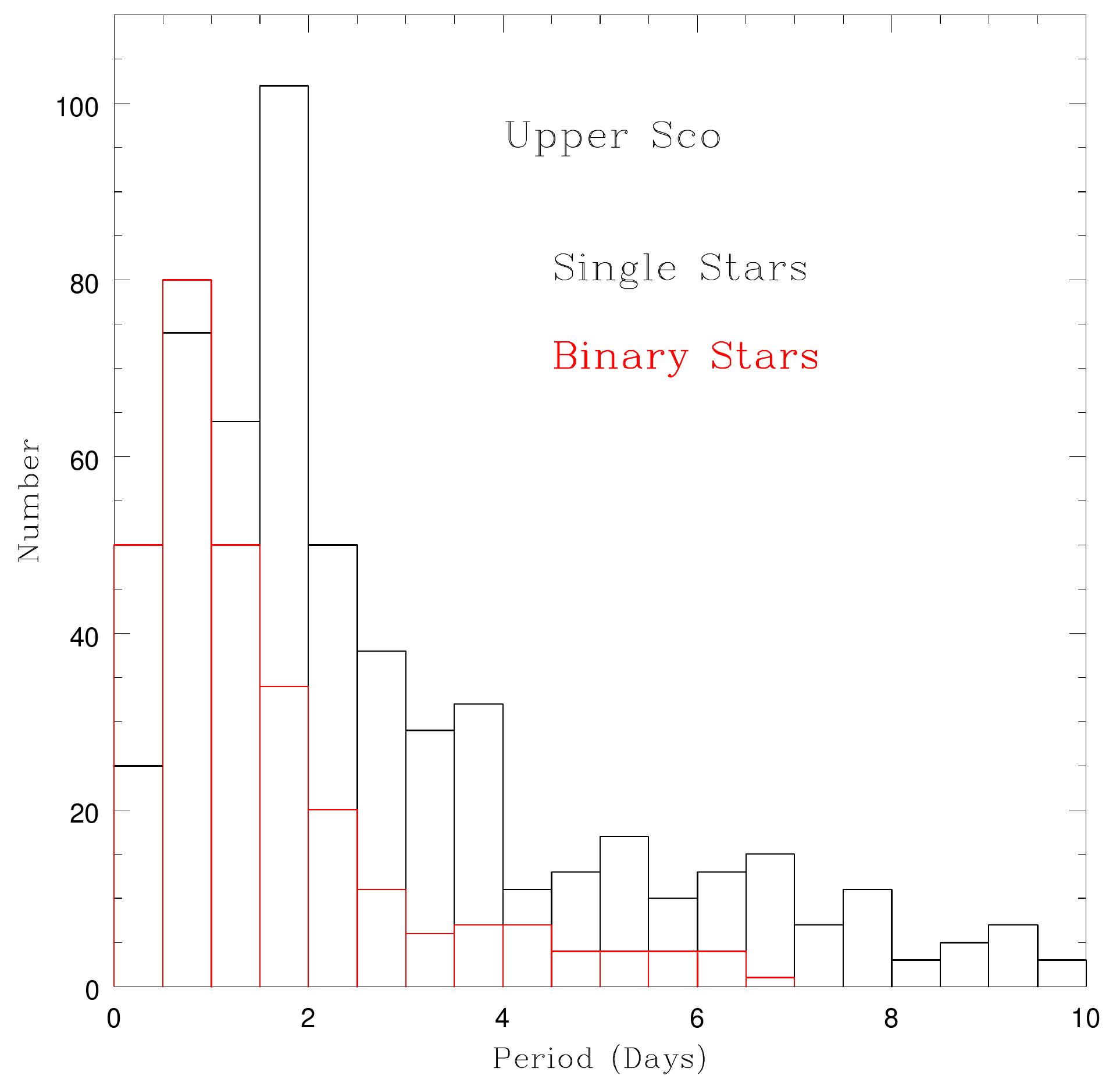}
\caption{Histogram of the rotation periods for both components of the
K2 binary stars with 4.0 $<(V-K_{\rm s})_0<$ 6.5 dM stars (roughly M0
to M5.5) in Upper Sco compared to the rotation periods of the single
dMs for the same color range. (black line= single stars; red line=
binary stars.)The binary stars are clearly more rapidly rotating.
\label{fig:Figure3}}
\end{figure}

As touched upon in Rebull \etal\ (2018a) and discussed more thoroughly
here, we have now found that for the entire dM color range where we
have good sampling (corresponding approximately to spectral types M0
to M5.5), the dM K2 binary stars  in the $\sim$8 Myr Upper Sco
association are also faster rotating and have rotation periods closer
to each other than if drawn from the single-star population -- with an
even greater level of significance than in the Pleiades. For this
color range (4.0 $<(V-K_{\rm s})_0<$ 6.5), we find 147 binary systems
and 562  single stars in Upper Sco with periods in our K2 sample.
Figure \ref{fig:Figure3} shows a histogram of the periods for the
components of the binary dM systems compared to the single stars; the
binary star members are clearly more rapidly rotating.   A KS test
shows that the probability that the binary and single star rotation
periods are drawn from the same parent population is less than 1 part
in one hundred thousand.   As another way to highlight the rapid
rotation of the dM binary stars, we note that among the 856 (562 +
2$\times$147) stars with periods in Figure 3, 34\% are members of
binaries; however, among the stars with $P <$ 1 day, 57\% are members
of binaries identified with K2 (i.e., with two rotation periods).

\begin{figure}[ht]
\epsscale{0.9}
\plotone{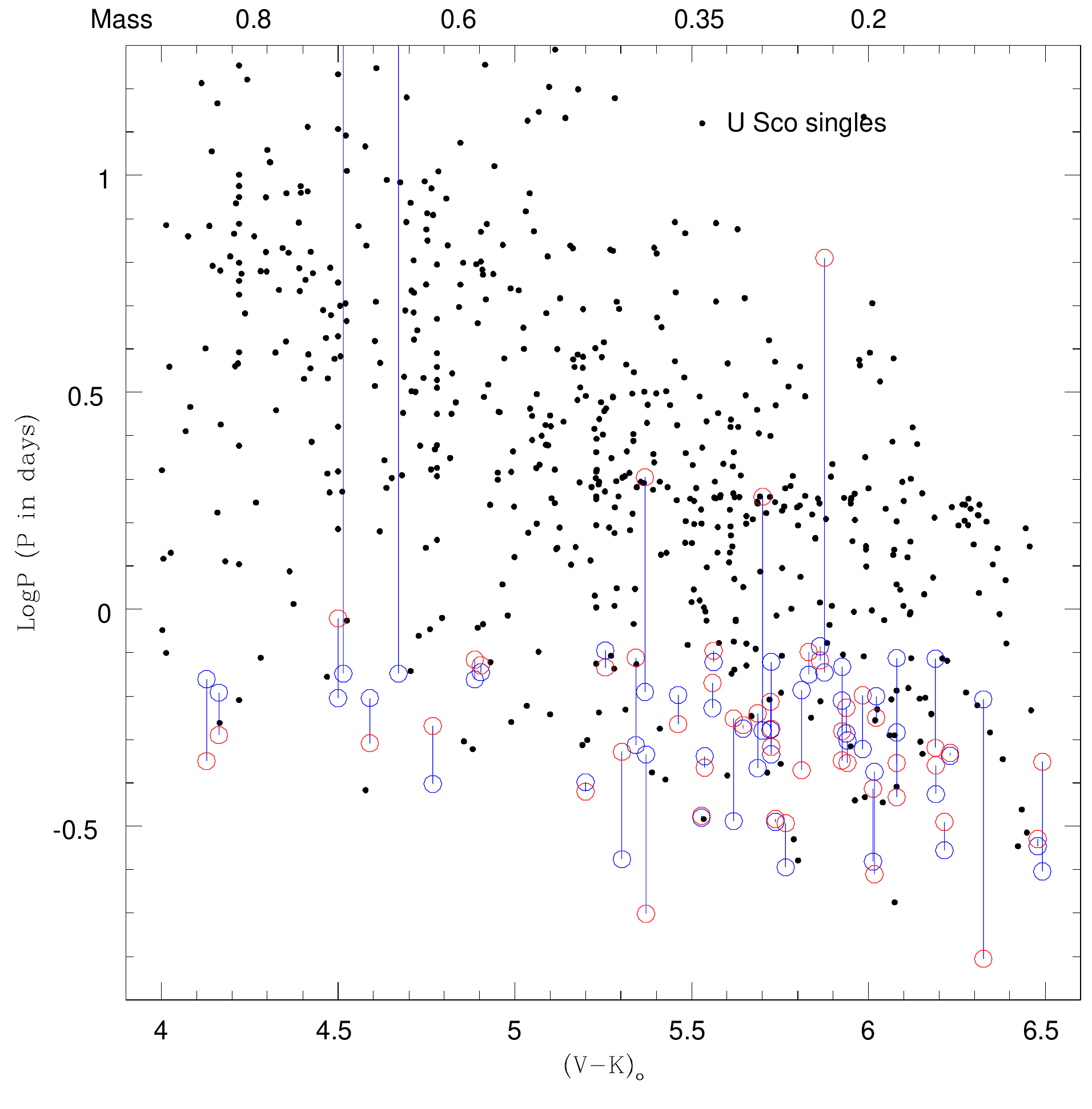}
\caption{Period-color plot for the Upper Sco dM stars.  Red and blue
circles connected by a bar are the periods for the two components of a
binary star; shown here are just the 50 most rapidly rotating (based
on the period of the primary star) binaries.  Black dots are the
periods of the stars with just one detected period in our K2 data.  
The primary star (blue circle) in each binary plotted here was chosen
to be very rapidly rotating -- but the secondary (red circle) could
have had a period anywhere within the region spanned by the black
points in the  diagram.  The fact that the secondaries nearly all have
periods quite close to that of the primary is a new and (perhaps)
unexpected result.
\label{fig:Figure4}}
\end{figure}

To illustrate the fact that the two components of the binary systems
have periods that are, on average, closer than expected to each other,
Figure \ref{fig:Figure4} provides a period-color plot where we
separately show (but link) the two components of the 50 binary systems
with the shortest primary-star periods and compare these systems to
the Upper Sco single dMs.  We plot only these most rapidly rotating
binary systems for clarity -- plotting the entire set of binaries
results in so many points in the diagram as to make it difficult for
the eye to pick out trends.   While the rapid rotation of the primary
stars in this plot is, in part, a selection effect, the fact that the
period of the secondary differs from that of the primary by an amount
so much less than the scatter in period at a given color for the
single stars is not a selection effect. In \S 4 and \S 5, we provide
more quantitative means to assess the significance of the difference
between the binary and single star rotation properties for late dM
stars in our four clusters.

\subsection{Assigning Masses and Adopting a Common Mass Range}

For the remainder of the paper, instead of sorting the stars by their
$(V-K_{\rm s})_0$ color, we will use their inferred stellar mass.  We
do this because the mass range for a given $(V-K_{\rm s})_0$ range in
Upper Sco is very different than the mass range for that same color
range in Praesepe/Hyades or the Pleiades.   While assigning masses for
dM stars on or very near the MS should be relatively  uncontroversial,
doing so at the age of Upper Sco is still a quite debatable process.  
In order to allow others to retrace our steps, we describe our process
for assigning masses to the stars for which we have K2 light curves in
the Appendix.

In order to compare the rotational data in the four clusters, we must
choose a common mass range.   As we noted previously, angular momentum
losses from winds eventually eliminates any initial spread in rotation
at a given mass and age (Skumanich 1972; Radick \etal\ 1987;
Pinsonneault, Kawaler \& Demarque 1990).  This is seen dramatically in
the K2 data for Praesepe (Rebull \etal\ 2017), where the great
majority of the FGK and early M dwarfs have a well-defined,
monotonically decreasing rotation rate with decreasing mass.  Any
initial spread in rotation rate at a given mass in that spectral type
range has been almost completely erased by $\sim$700 Myr.  We
therefore choose to set the upper mass limit for our study at the
point in mass where the Praesepe slow sequence ends -- which is at
about 0.32 \msun.  This upper mass limit also places us in the mass
regime where it is safe to assume that two peaks in the LS periodogram
identifies the star as a binary (see Figure 2).   For the lower mass
limit, we choose 0.1 \msun,  because it corresponds approximately to
the faint limit where we have reasonably good sensitivity and are able
to determine periods in most of the stars with K2 data. 

With those choices and with the other demonstrations in the preceding
sections, we are now set to proceed to compare the rotation rates of
the binary and single stars in our four clusters.

\section{The Correlation between Rotation and Binarity for Fully-Convective Low Mass Stars at Early Ages}

\begin{figure}[ht]
\epsscale{0.9}
\plotone{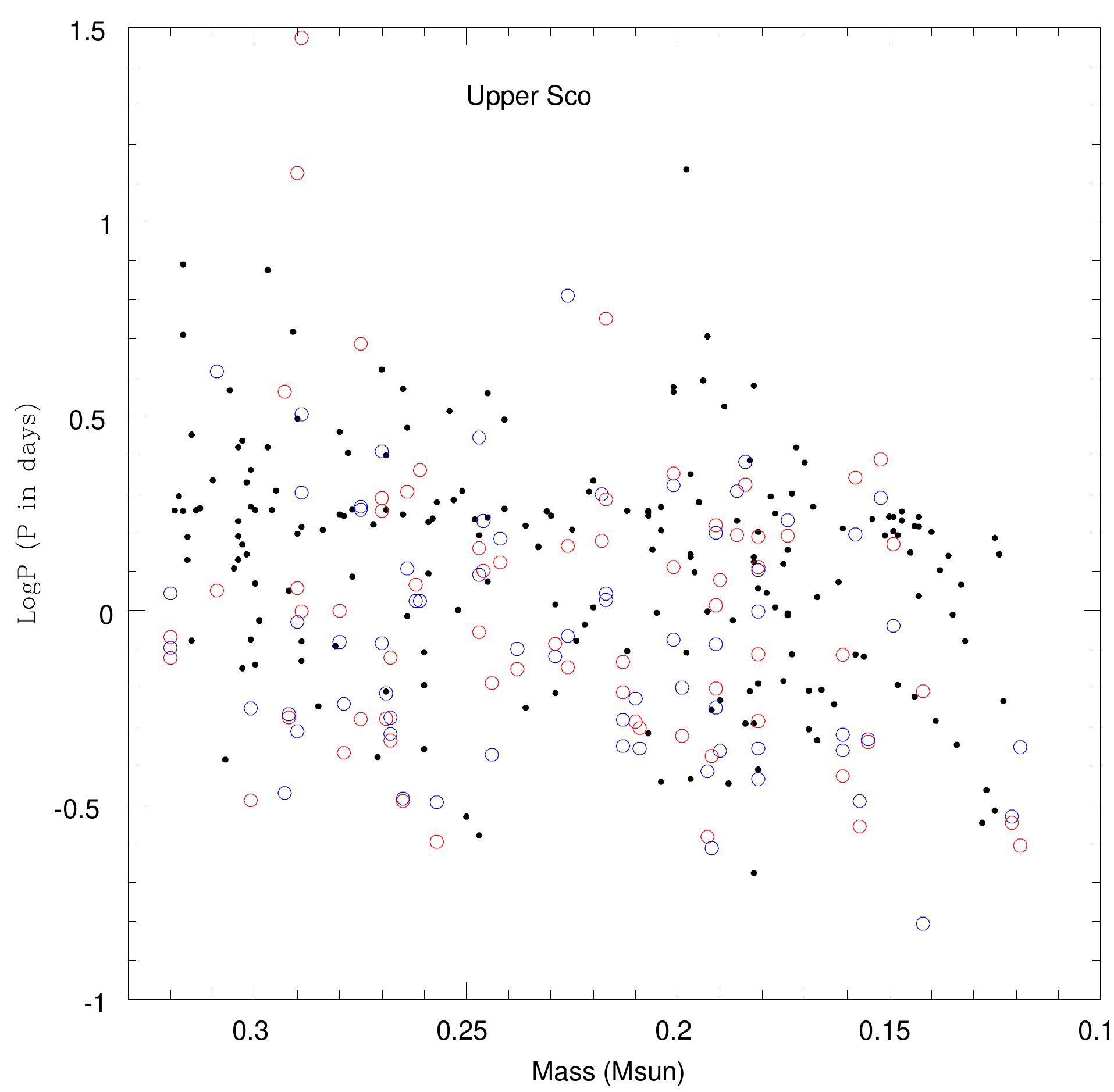}
\caption{The correlation between rotation and mass for Upper Sco M
dwarfs with $M \leq$ 0.32 \msun\ and  with K2 rotation periods. The
blue and red circles correspond, respectively, to the periods of the 
primary and secondary components of the K2 binaries; the black dots
mark periods for the stars identified as single by their K2 data.  The
binary stars identified with the K2 data dominate the most rapidly
rotating part of the distribution, while single stars dominate the
most slowly rotating region of the diagram.
\label{fig:Figure5}}
\end{figure}

\begin{figure}[ht]
\epsscale{0.9}
\plotone{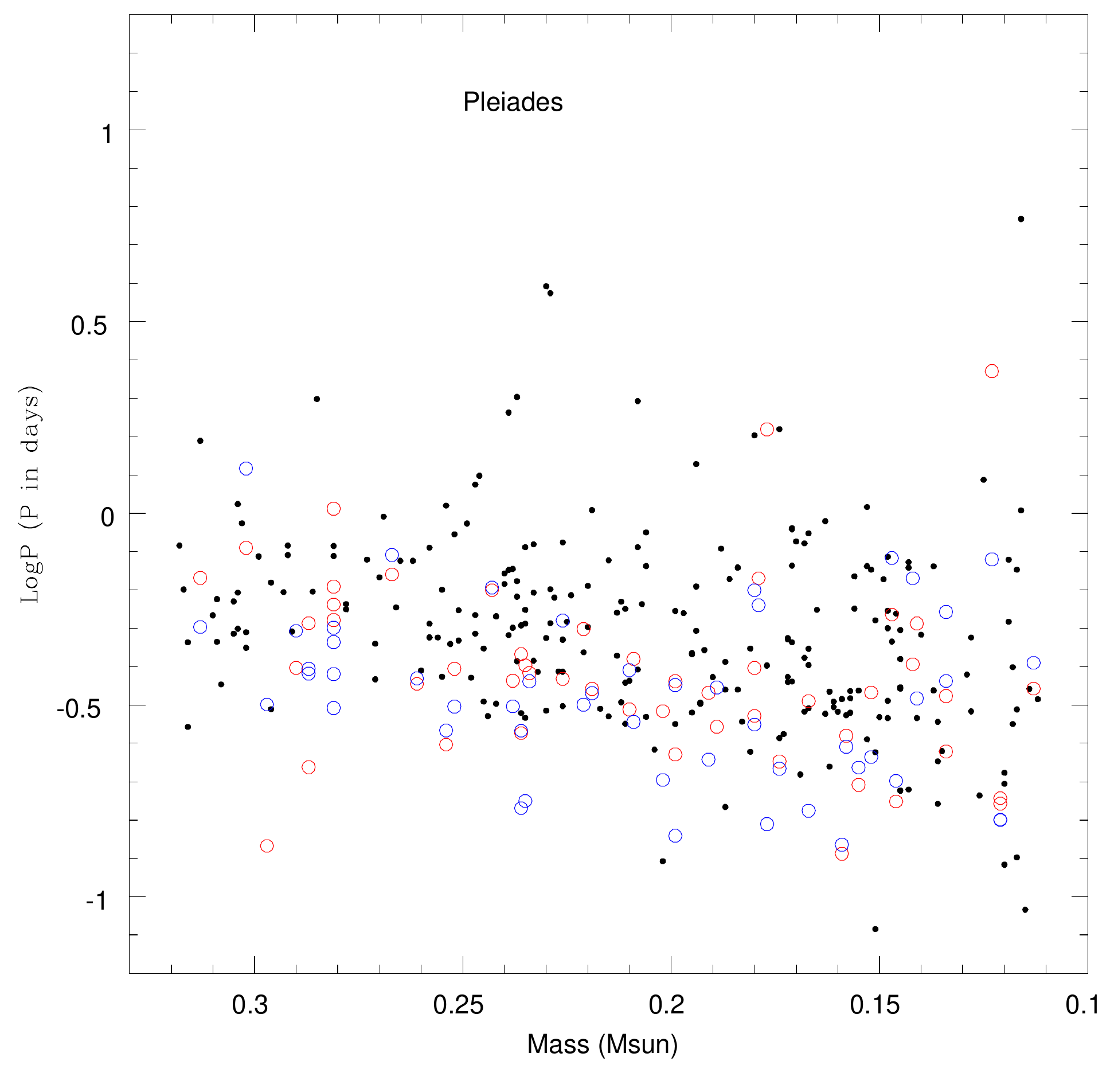}
\caption{The  correlation between rotation and mass for Pleiades  M
dwarfs with $M \leq$ 0.32 \msun\ and  with K2 rotation periods.  The
stars that are components of binary stars identified with the K2 data
dominate the most rapidly rotating part of the distribution. (Notation
is as in prior figure.)
\label{fig:Figure6}}
\end{figure}

\begin{figure}[ht]
\epsscale{0.9}
\plotone{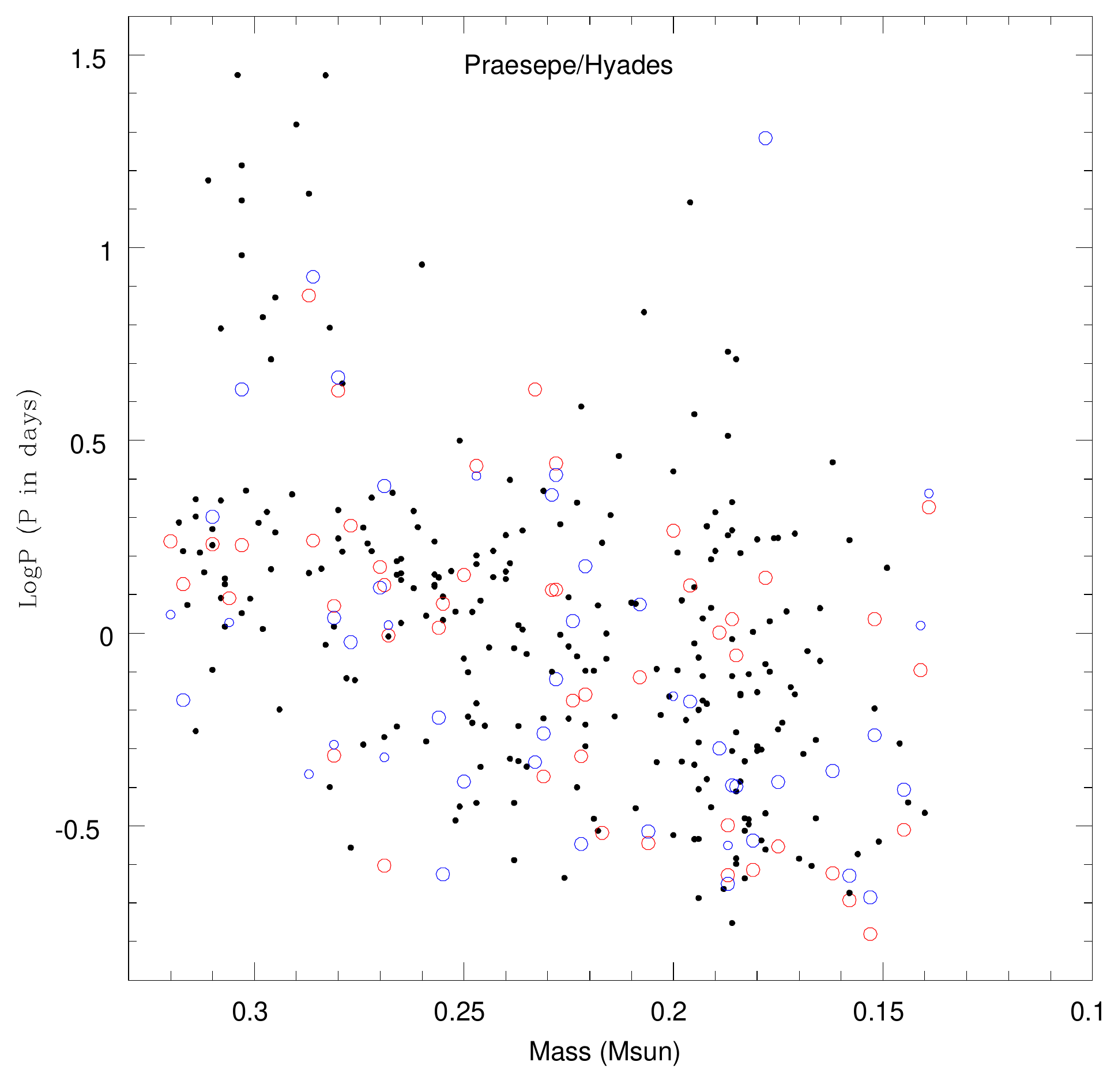}
\caption{The  correlation between rotation and mass for
Praesepe/Hyades M dwarfs with $M \leq$ 0.32 \msun\ and  with K2
rotation periods.  The stars that are components of binary stars and
the single stars do not obviously have different distributions in this
diagram. (Notation is as in prior figure.)
\label{fig:Figure7}}
\end{figure}

Using the mass calibrations we describe in the Appendix, adopting a
mass range of 0.1 $< M$/ \msun\ $<$ 0.32, we have a total sample of 257
stars (191 with one K2 period and 66 with two periods) in Upper Sco,
286 stars in the Pleiades (236 with one period, 50 with two) and 292
stars in Praesepe/Hyades (245 one period, 47 two periods).  Tables
listing these stars, their \vmk\ colors, rotation periods and masses
are provided in the Appendix.   Figures \ref{fig:Figure5},
\ref{fig:Figure6},  and \ref{fig:Figure7}  show mass-period diagrams
for the four clusters.  Each K2 binary appears twice in these figures,
with the component having the stronger periodogram peak in blue and
the star with the  weaker periodogram peak in red; stars with just one
K2 period are shown as black dots. While perhaps not ``obvious",
careful visual inspection of these figures suggests that in this mass
range, the binary stars rotate faster, on average, than the single
stars in Upper Sco and probably in the Pleiades.  

\begin{figure}[ht]
\epsscale{0.9}
\plotone{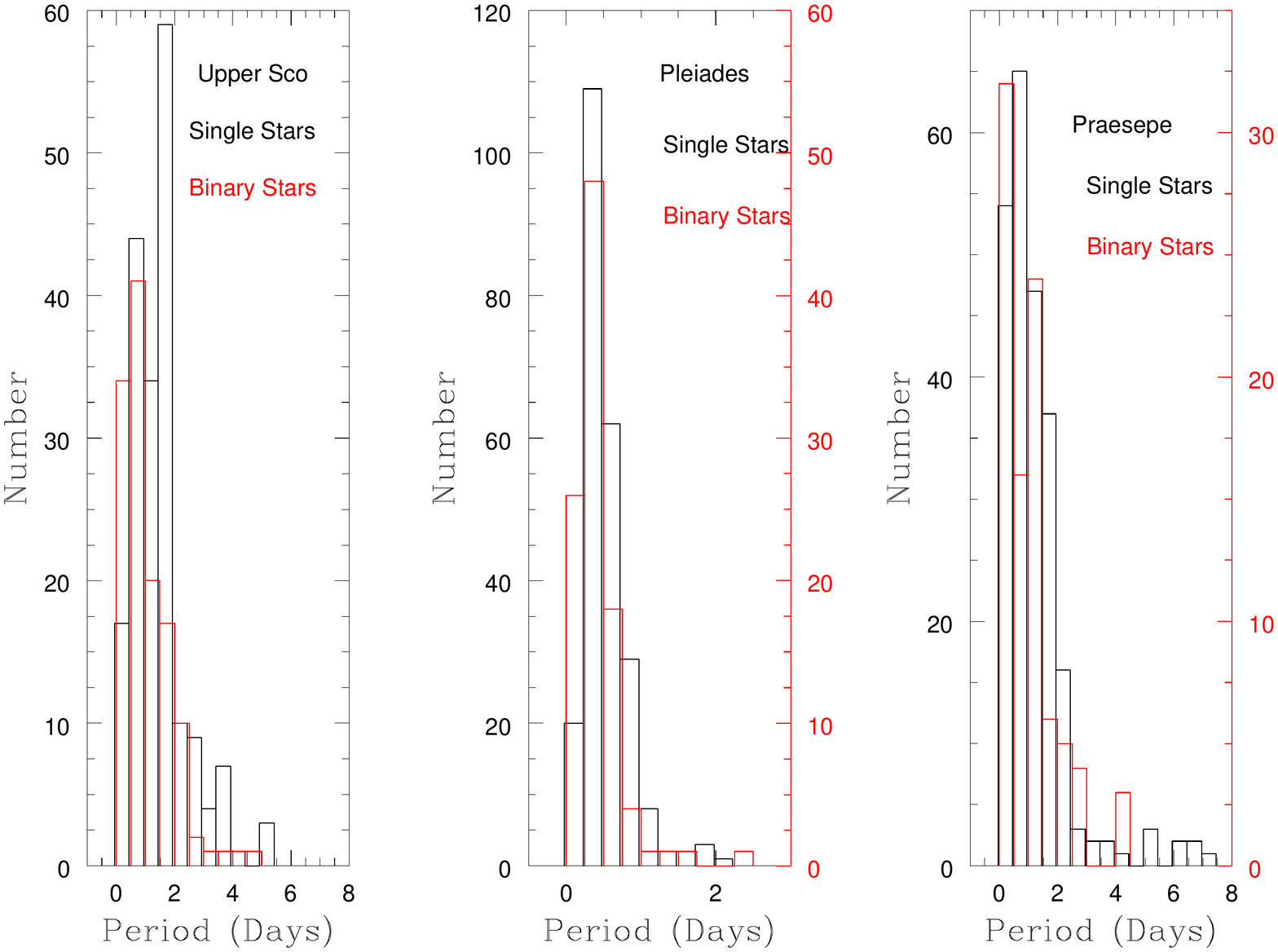}
\caption{For stars with inferred masses $<$ 0.32 \msun, histogram
comparison of  the rotation periods for K2 binary stars vs.\ stars with
only one rotation period in the K2 data (hence assumed to be single or
a low-$q$ binary).  At young ages, the binary stars are weighted towards
rapid rotation compared to the single stars.   At Praesepe age, the
distributions for the singles and binaries appear, by eye,  to be more
similar.   For the Pleiades and Praesepe, separate $y$-axis scales are
used for the singles (left axis) and binaries (right axis) because there
are only half as many binary star periods as there are single star 
periods.
\label{fig:Figure8}}
\end{figure}

Collapsing the distributions into period histograms makes the
difference between the single stars and the binaries somewhat more
obvious.  Histograms for the four clusters are provided in Figure
\ref{fig:Figure8}.   The Upper Sco histogram now clearly reveals that
the binary stars are more weighted to short periods than the stars
with just one K2 period.   The histogram for binaries and singles in
Praesepe/Hyades, however, shows that those two distributions are very
similar.   These qualitative conclusions are confirmed by applying the
KS test to the rotation data for the  $M <$ 0.32 \msun\ dM stars in
the three clusters.   That test reveals that the probability that the
binary and single star rotation periods in Upper Sco are drawn from
the same parent population is less than 1 part in 10$^5$; for the
Pleiades, the probability is less than 1 part in 10$^4$.  However, for
Praesepe/Hyades, the KS test yields $p \sim$ 0.12, indicating that the
singles and binaries are consistent with being drawn from the same
parent population\footnote{There is a strong correlation between mass
or $(V-K_{\rm s})_0$ color and period in these four clusters,
potentially affecting these KS tests.   As long as the distribution in
color or mass for the binary and single stars is about the same, which
is indeed the case, this should not greatly affect the KS test
results.   However, to be certain, we have also fit and removed the
trend of median period as a function of color from the data for each
cluster and repeated the KS test analyses.   In all cases, these KS
test probabilities confirm the original ones, with the same or higher
degrees of significance.}.  

Another way to emphasize the difference in rotation rates of binary
and single stars in the three populations is to isolate the most
rapidly rotating and most slowly rotating stars and ask what fraction
of those stars are in binaries and what fraction are in singles.  In
Upper Sco, 41\% (132/323) of all the $M <$ 0.32 \msun\ stars are in K2
binaries; however, among the 20\% most rapidly rotating of the whole
sample, 67\% are in K2 binaries whereas among the 20\% most slowly
rotating stars, 36\% are in K2 binaries.  In the Pleiades, 30\% of all
the late dM stars are in K2 binaries, but 49\% of the 20\% most
rapidly rotating stars are in K2 binaries and 12\% of the 20\% most
slowly rotating stars are in K2 binaries.   Finally, for
Praesepe/Hyades, 28\% of all the stars are in K2 binaries, while 38\%
of the 20\% most rapidly rotating stars are in K2 binaries and 25\% of
the 20\% most slowly rotating stars are in K2 binaries.  These data
show that binary stars are over represented among the most rapidly
rotating stars  in each cluster and under represented among the most
slowly rotating stars - and that this bias is strongest at young ages.

Thus, assuming that the stars in these four clusters can be viewed as
different snapshots in time of a single population, the rotation
periods of the $M <$ 0.32 \msun\ binaries are faster at very young
ages than their single cousins but angular momentum loss mechanisms
apparently drive those rotation distributions to become more similar
over time.

\section{Monte Carlo Simulations and Other Statistical Measures}

\begin{figure}[ht]
\epsscale{0.9}
\plotone{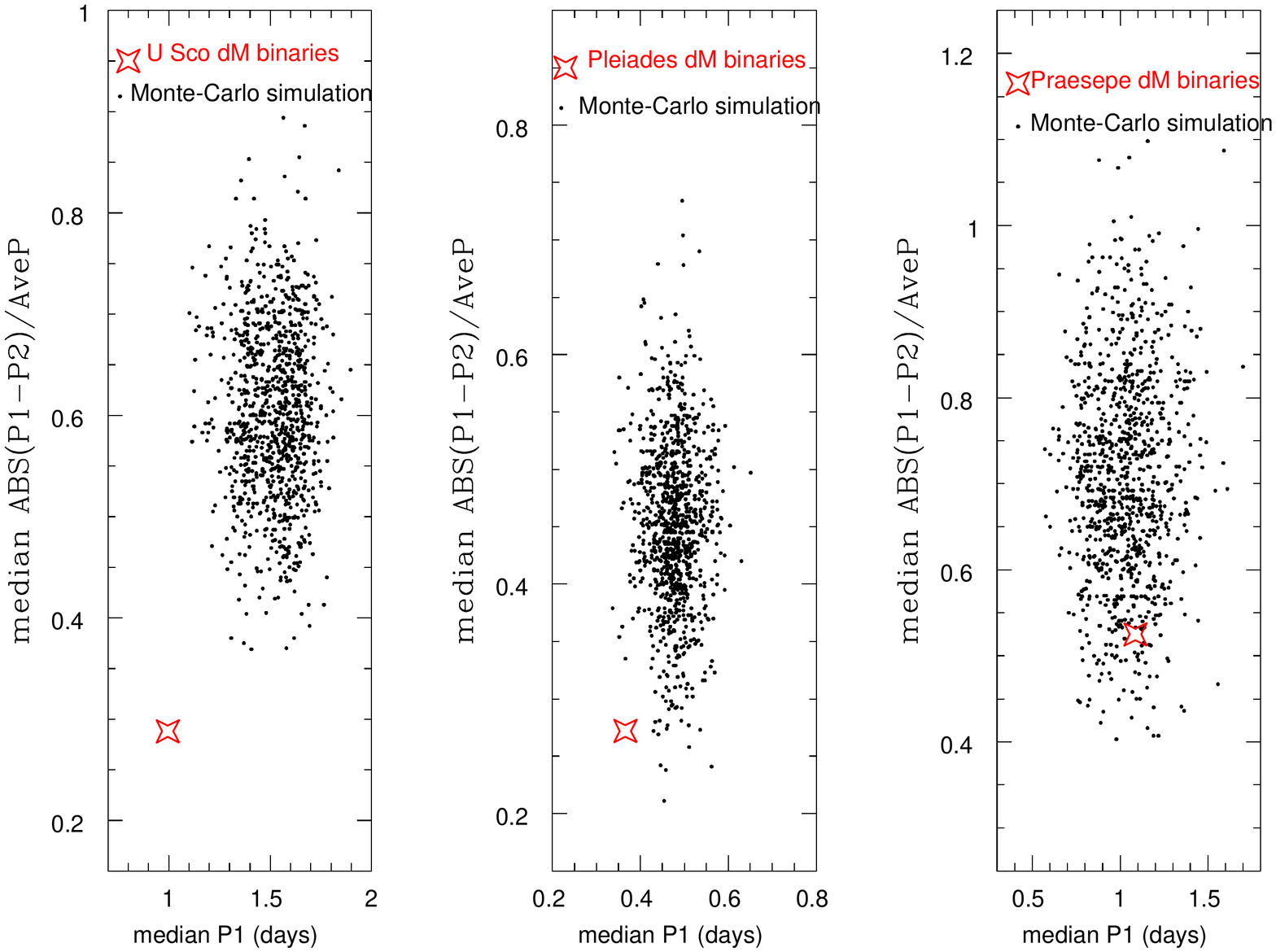}
\caption{Comparison of the median $P_1$ and median normalized period
difference for the actual K2 dM binaries in our four clusters versus
the same data for simulated binary populations drawn at random from
the single star periods.  Each black dot represents the result of one
simulation; each simulation creates the same number of binary stars as
we have found in our K2 data for that cluster, with the two periods
drawn at random from the single star population whose $(V-K_{\rm
s})_0$ colors are within 0.15 mag of one of our binaries. We ran 1000
simulations for each cluster (resulting in the 1000 black dots for
each cluster in the figure).
\label{fig:Figure9}}
\end{figure}

In order to more fully compare the rotation periods of the late dM
binaries in our four clusters to the single dMs, we have created a
Monte Carlo simulation routine.  For each run, the routine produces a
simulated set of binary star periods drawn from the periods of the
single stars, with the number of simulated binaries being the same as
the actual number of binaries in that cluster.  Because the companions
we detect with K2 must have magnitudes that are relatively similar to
that of the primary, for each simulated binary  we restrict the second
period to come from a star within 0.15 magnitude in $(V-K_{\rm s})_0$
from the star selected randomly as the primary.  For each binary star,
we calculate a normalized period difference defined as $dP_n =
|P_1-P_2|/[(P_1+P_2)/2]$.   For the ensemble of simulated binaries in
a given run, we derive the median normalized period difference and the
median period for the primary star.   We run 1000 such simulations for
each cluster.  Figure \ref{fig:Figure9} compares the median normalized
period and median first period for the actual set of binaries in each
cluster to the median results for each of the 1000 simulations. The
red star symbols in each of the plots denote the median primary period
and median normalized period difference for the actual dM binaries in
each cluster.   The size of those symbols was chosen to provide a
rough estimate of the statistical uncertainty in those values. 
Because the distributions of period and the normalized period
differences are not Gaussian in shape, we cannot estimate those
uncertainties as simply the standard deviation about the mean divided
by the square root of the number of binary stars.  Instead, we define
$\pm 1 \sigma$ as the range about the median in period (or normalized
period difference) within which 67\% of the values fall, and then we
assume the uncertainty in the median scales with the square root of
the number of points.

The simulations confirm that the actual mid-dM K2 binaries in Upper
Sco and the Pleiades have periods that are shorter and closer to each
other than the simulated populations drawn from the single stars.  
The Praesepe/Hyades mid-dM K2 binaries have periods consistent with
having been drawn from their single star population.   

\section{The Influence of Disks}

In standard models of the angular momentum evolution of low mass
stars, a common way to explain the presence of both slowly rotating
and rapidly rotating populations at young ages is to include the
influence of primordial circumstellar disks on stellar rotation
(K\"onigl 1991; Bouvier \etal\ 1997; Sills \etal\ 2000).   When the
disks are still present and actively accreting, they can drain angular
momentum from the star and keep it slowly rotating.  This could be
tied to a difference in the rotation rate of single and binary stars
if the binary stars are close enough to each other to disrupt or
truncate a circumstellar disk that might be present around one or both
components of the binary.   In Rebull \etal\ (2018a) we presented some
direct evidence that ``disk-locking" is at work in at least a few of
the Upper Sco stars with disks (classical T~Tauris, CTTs), and that
the low mass CTTs do rotate more slowly than the WTTs stars without
IR excesses.   Here we provide additional evidence that
disks seem to have a particularly strong and deterministic influence
on the rotation rates of $M <$ 0.32 \msun\ stars in Upper Sco.

\begin{figure}[ht]
\epsscale{0.9}
\plotone{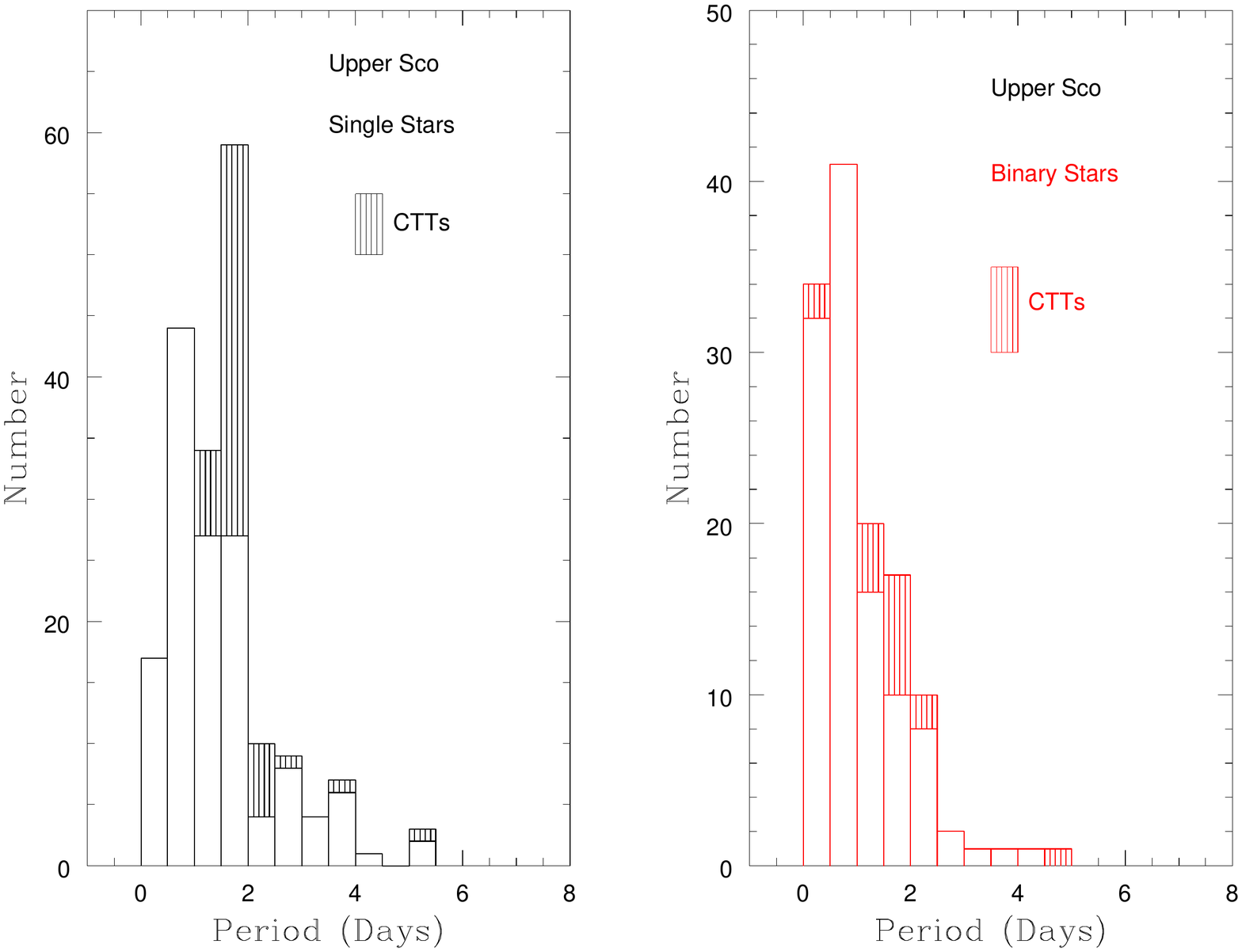}
\caption{Same data as plotted in Figure 8 for Upper Sco, except now
marking the stars which we believe to have IR excesses (with vertical hash
marks), and therefore likely to have on-going accretion.   The left
hand panel shows the rotational period distribution for single
mid-dMs; the right hand panel shows the distribution for both
components of the K2 binary mid-dMs (there are eight binary systems
with IR excess; thus sixteen of the periods in the right-hand-side
plot are associated with CTTs). A large part of the difference between
the binary and single star rotational distribution is due to the
slowly rotating population of stars with IR excesses.
\label{fig:Figure10}}
\end{figure}

The rotation period histogram for Upper Sco shown in the left-hand
panel of Figure \ref{fig:Figure8} includes all members in that mass
range.  Rebull \etal\ (2018a) identified 56 of those stars as
possessing IR excesses consistent with primordial disks and hence
their being CTTs.  In Figure \ref{fig:Figure10}, we replot the Upper
Sco rotation period histograms for single and binary mid-dM stars, but
this time also marking those stars that have IR excesses.  The disked
stars have periods strongly peaking around 1-2 days.    Comparing only
the stars without IR excesses, the histograms for the single and
binary stars are significantly more similar; a KS test now yields $P
\sim$ 0.059,  indicating these two distributions are consistent
(though only marginally)  with having been drawn from the same parent
population.

\begin{figure}[ht]
\epsscale{0.9}
\plotone{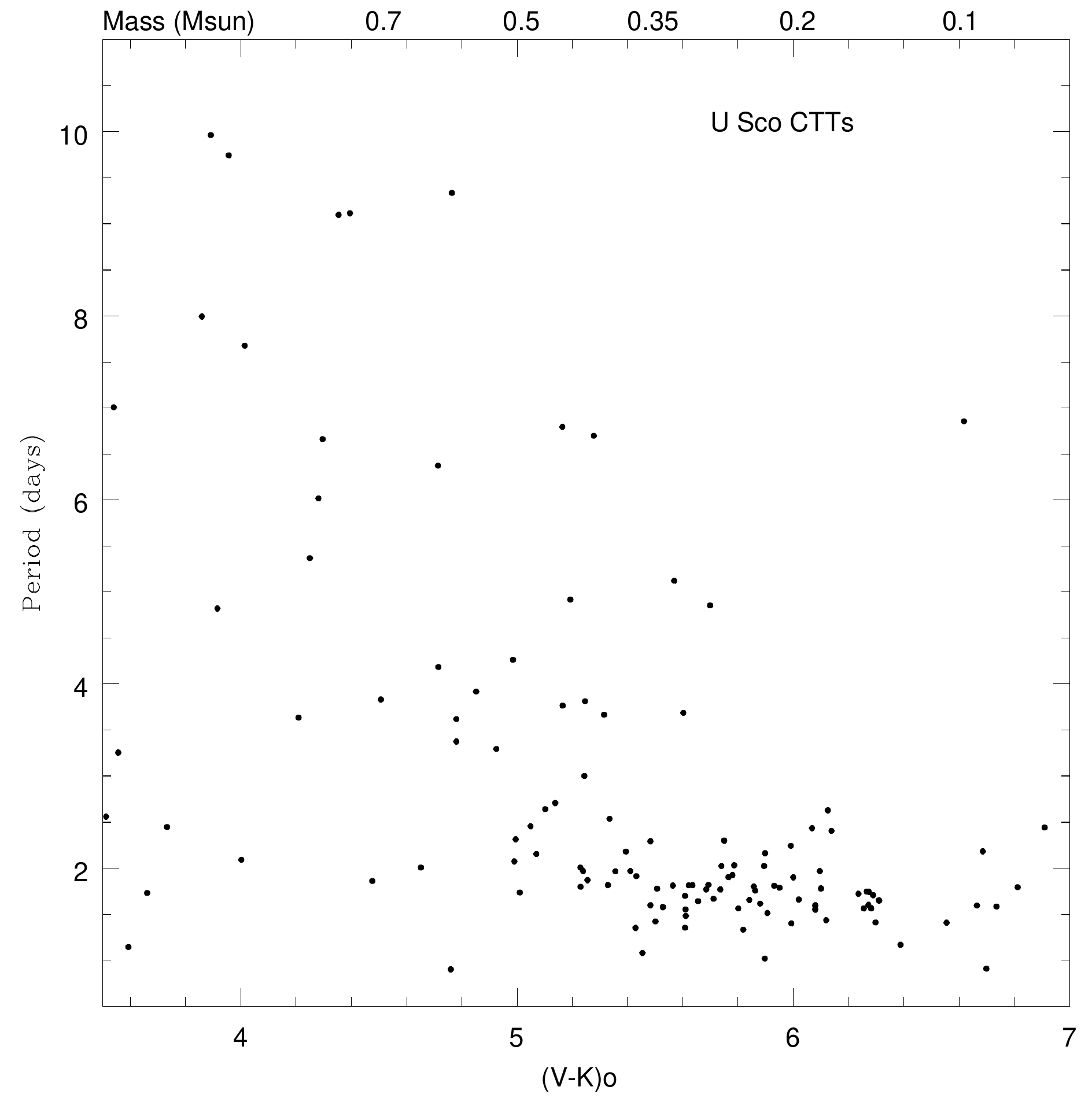}
\caption{Rotation periods for just the Upper Sco CTTs for which we have K2
data.  $M =$ 0.32 \msun\ corresponds approximately to $(V-K_{\rm s})_0$= 5.3.  Redward
of that color, the great majority of the CTTs have periods between 1 and 2 days;
blueward of that color, there is a much wider range of rotation periods. 
\label{fig:Figure11}}
\end{figure}

Figure \ref{fig:Figure11} shows these same data in another way, also
intended to highlight the link between IR excesses and rotation for
very low mass stars in Upper Sco as well as  the strikingly narrow
range in periods for the $M <$ 0.32 \msun\ Upper Sco CTTs.    In this
plot of period versus $(V-K_{\rm s})_0$ color for just the Upper Sco
CTTs, there appears to be a sharp transition at $\sim$0.4 \msun\
($(V-K_{\rm s})_0\sim3$), with a wide range in rotation periods for
higher masses and a quite narrow range in rotation periods below that
mass.  If this is due to magnetohydrodynamic processes linking the
star and the inner disk, then plausibly there is some change in either
disk properties or magnetic field properties at that mass boundary.  
We have looked at the SEDs of the CTTs above and below the boundary
mass, and do not see any systematic differences (there are a range of
SED shapes from transition disk to full Class II in both mass
ranges).   This could, therefore, be a signature of a change in
magnetic field topology near this mass at Upper Sco age.

\section{Additional Thoughts re: Binarity and Rotation in Upper Sco and the Pleiades}

If binarity links to rotation by suppressing star-disk angular
momentum regulation, then knowledge of the orbital separation between
the components of the binary becomes an important topic.   
Unfortunately, the $M <$ 0.32 \msun\ stars in even these relatively
nearby clusters are quite faint and there is little published
literature on their orbital periods/separations.  However, if one
opens up the topic to all dM members, then there is more published
literature, particularly for Upper Sco where its youth makes the M
stars comparatively bright.

The two most extensive and sensitive high spatial resolution imaging
studies of Upper Sco (Kraus \etal\ 2008; Lafreniere \etal\ 2014)
include 20 spectral type K7 or later members with K2 data.  One
additional M2.5 K2 member was observed with Keck LGS-AO by Ansdell
\etal\ (2016).  The pertinent data for these 21 stars are provided in
Table \ref{tab:USCO_AOdata}.  The reddest of these stars has
$(V-K_{\rm s})_0$ = 5.46, corresponding to essentially $M$ = 0.32
\msun, reinforcing the claim that for $M <$ 0.32 \msun, there is
little published high spatial resolution imaging data.   However, for
earlier type dM members of Upper Sco, the data in Table
\ref{tab:USCO_AOdata} indicate that:
\begin{itemize}
\item None of the AO binaries with $\Delta K >$ 1.0 mag are detected
 with two periods by K2, in agreement with the belief that K2 is
 detecting primarily ``high-$q$" systems.
\item Four of the five AO binaries with $\Delta K <$ 0.2 mag do have two periods
  detected by K2, also supporting that notion.
\item Essentially half of the AO binaries are also K2 binaries.
\item Nine of the 10 K2 binaries have $\Delta K \leq$ 0.6 mag.  Eight
   of the K2 binaries have angular separation $<$ 0.33$\arcsec$
   (projected spatial separation $<$ 45 AU).
\end{itemize}

\floattable
\begin{deluxetable*}{lcccccc}
\tabletypesize{\footnotesize}
\tablecolumns{7}
\tablewidth{0pt}
\tablecaption{Upper Sco dM Binaries with Both AO and K2 data\label{tab:USCO_AOdata}}
\tablehead{
\colhead{EPIC Number} &
\colhead{Spectral Type } &
\colhead{$P_1$} &
\colhead{$P_2$} &
\colhead{$\Delta K_{\rm s}$\tablenotemark{a}} &
\colhead{Separation\tablenotemark{a}} &
\colhead{Name}  \\
 & & \colhead{(days)}& \colhead{days}
&  \colhead{(mag)} & \colhead{arcsec}  &  }
\startdata
205089832 &  M2 &  8.1750 &  \nodata &  1.59 &  0.092 &  USco 160707.7-192715 \\
205060410 &  M1 &  17.9258 &  \nodata &  0.98 &  0.652 &  USco 160823.8-193551 \\
205137430 &  K7 &  12.2019 &  \nodata &  3.83 &  5.775 &  USco 161031.9-191305 \\
204251947 &  M1 &  5.4849 &  \nodata &  0.37 &  1.981 &  GSC 06793-00868 \\
204342099 &  M1 &  \nodata & \nodata  &  1.19 &  1.907 &  GSC 06793-00806 \\
204179058 &  M1 &  3.0039 &  0.5857 &  0.78 &  0.054 &  ScoPMS 017 \\
204436170 &  M1 &  0.8621 &  0.7271 &  0.03 &  0.025 &  ScoPMS 019 \\
205167008 &  M3 &  4.1422 &  \nodata &  2.48 &  4.61 &  ScoPMS 042b \\
204878974 &  M1 &  3.0938 &  0.8548 &  0.43 &  0.189 &  RX J1600.5-2027 \\
204794876 &  M0 &  1.4900 &  2.1528 &  0.58 &  0.205 &  RX J1601.7-2049 \\
203834337 &  K7 &  4.6547 &  \nodata &  1.00 &  0.076 &  RX J1601.8-2445 \\
204894575 &  K7 &  1.9540 &  \nodata &  0.18 &  0.310 &  RX J1602.9-2022 \\
204862109 &  M0 &  1.7281 &  1.0520 &  0.53 &  0.121 &  RX J1603.9-2031B \\
205142483 &  M1 &  6.6342 &  \nodata &  1.47 &  0.599 &  RX J1607.0-1911 \\
204040060 &  M0.5 &  7.2390 &  \nodata &  0.63 &  1.324 &  ScoPMS 016 \\
204406748 &  M3 &  4.3922 &  16.9554 &  0.60 &  0.193 &  ScoPMS 020 \\
205164892 &  M1 &  6.6595 &  \nodata &  0.42 &  0.299 &  ScoPMS 042a \\
203895983 &  M2.5 &  2.4474 &  2.5757 &  0.1 &  0.298 &    RIK 77 \\
203628765 &  M0 &  0.5034 &  0.6615 &  0.07 &  0.594 &  [PZ99] J155716.6-252918 \\
205087483 &  M4 &  1.5134 &  2.3752 &  0.05 &  0.862 &  [PGZ2001] J160801.5-192757 \\
205374937 &  M4 &  0.6345 &  0.5436 &  0.05 &  0.098 &  [PGZ2001] J161118.1-175728 \\
\enddata
\tablenotetext{a}{The $\Delta K$ and separation data are from Kraus \etal\ (2008),
Lafreniere \etal\ (2014), and Ansdell \etal (2016).}
\end{deluxetable*}
\noindent

The AO data are therefore generally supportive of the idea that our K2
dM binaries in Upper Sco (and presumably Pleiades and Praesepe) are
high-$q$, relatively small separation binaries.   However, because of
the very poor overlap with our $M <$ 0.32 \msun\ primary dataset, any
more definitive conclusion must await deeper, higher spatial
resolution data.

The other avenue we can pursue with existing data is to ask ``Why do
we not detect two periods for all of the photometric binaries that lie
well above the single-star locus in the optical CMDs for these
clusters (see Figure 2)?" Perhaps our K2 binaries are biased to short
periods (relative to the full set of binaries), and the correlation we
see between binarity and rotation is thus illusory.   We cannot
address this issue with Upper Sco data because the uncertainties in
the dereddened colors are too large (specifically, we cannot identify
a reliable photometric binary sample); therefore, we will use the
Pleiades data.   We make these tests for the full set of Pleiades dM
stars (4.0 $<$ \vmk\ $<$ 6.5) for which we have K2 data in order
to have as large a sample size as possible.  First, we have compared
the rotation periods of the photometric (but not K2) dM binaries with
$\Delta V >$ 0.4 mag to the dominant rotation periods of the K2 dM
binaries.  Those two distributions are very similar; a KS test yields
$p$ = 0.42, indicating that the two distributions are consistent with
having been drawn from the same parent population.  Next, we compared
the rotation periods of those same photometric (but not K2) binaries
to the rotation periods of the dM stars lying within 0.25 mag of the
single-star locus in the $V$ vs.\ \vmk\ CMD.   That KS test
yields $p$ = 0.0159, indicating the two sets of stars are not drawn
from the same parent population, with the K2 single but photometric
binary stars rotating faster than the photometric single stars.   We
conclude from this that the dM stars in the Pleiades that are
photometric binaries but where we  detect just one K2 period are also
relatively rapidly rotating.  We suspect that the reason we do not
detect a second period in these systems is not due to any single
cause, but could include secondaries that are nearly pole-on, or stars
that just happen to have relatively axisymmetric spot distributions,
or systems where the two periods are so close together we cannot
separate them in the LS periodogram, or where the quality of the K2
data are poorer than usual.   We discuss these topics at length in the
last section of the Appendix.

\section{Conclusions and Path Forward}

We have used the excellent photometric time-series light curves from
NASA's K2 mission to ask the question, ``Are the rotation rates of the
components of very low mass binary stars different from that of their
similar mass single stars at  young ages?"  The answer to that
question is yes, at least for photometric binaries.  
Late-type dM stars in Upper Sco and the Pleiades
with two detected periods (which we call ``K2 binaries") have rotation
periods distinctly shorter than single stars of that mass.  We have
also shown that the rotation periods of the two stars in these
binaries are significantly more similar to each other than would be
true if the periods were drawn at random from the single star
population.  These facts could in principle be due to at-formation
processes whereby the components of the binary stars began their lives
with more angular momentum than single stars.   Or, it could result
from angular momentum regulation mechanisms at young ages that might
be influenced by the presence or absence of a companion star.   We
believe the latter explanation is more likely, based on our
demonstration that much of the difference in the rotation period
distributions for the single and binary dM stars in Upper Sco is due
to a population of (relatively) slowly rotating CTTs.     A plausible
physical explanation of the slow rotation of the single dM stars is,
therefore, that those stars retained their primordial circumstellar
disks for a longer period and that disk-locking (K\"onigl 1991) or some
similar process allowed them to drain away some of their angular
momentum.   The binary dM stars disrupt their disks and thus spin up
more during PMS contraction.

By comparing populations of low mass dM stars at three ages (8 Myr,
125 Myr and 700 Myr), we have shown that the difference in rotational
properties between the single and binary stars decreases with
increasing age.  This must arise as a result of the detailed
dependence of angular momentum loss from winds on rotation rate for
stars in this mass range (0.1 to 0.3 \msun).   We intend to examine
this topic in more detail in a future paper (Pinsonneault \etal\
2019).

We note that our detection method probes only binaries with masses
close  enough for us to be able to measure the rotation periods of
both components.   Higher mass ratio binaries could in principle have
different rotation  distributions, and it will be difficult to infer
the spin rates of the  secondaries in such systems. Our working
hypothesis is that the difference  in rotation rates is more logically
tied to orbital period than to mass  ratio, and that we see an effect
because binaries with low mass ratios  are more likely to be close. A
testable prediction would therefore be  that we would expect wider
binaries to have rotation rates similar to single stars. The most
important next step in order to use the very low mass stars in these
clusters to help us better understand the formation and evolution of
single and binary stars will be to obtain high spatial resolution
imaging and  multi-epoch radial velocity data in order  to determine
or constrain the separations (or orbital periods) and mass ratios of
the binary stars in our four clusters -- with the most important
targets being the younger (Upper Sco and Pleiades) stars.   In Upper
Sco, in particular, it may be possible to use Gaia DR2 data to
determine a better member list and to more accurately plot stars in a
CMD by providing accurate, individual distances.  Spectra of
sufficient quality to determine quantitative spectral indices could
provide a better means to produce an equivalent CMD and (with Gaia DR2
distances) thereby  yield a photometric binary sequence of similar
quality as for the older clusters.

\begin{acknowledgements}

Some of the data presented in this paper were obtained from the
Mikulski Archive for Space Telescopes (MAST). Support for MAST for
non-HST data is provided by the NASA Office of Space Science via grant
NNX09AF08G and by other grants and contracts. This paper includes data
collected by the Kepler mission. Funding for the Kepler mission is
provided by the NASA Science Mission directorate. This research has
made use of the NASA/IPAC Infrared Science Archive (IRSA), which is
operated by the Jet Propulsion Laboratory, California Institute of
Technology, under contract with the National Aeronautics and Space
Administration. This research has made use of NASA's Astrophysics Data
System (ADS) Abstract Service, and of the SIMBAD database, operated at
CDS, Strasbourg, France. This research has made use of data products
from the Two Micron All-Sky Survey (2MASS), which is a joint project
of the University of Massachusetts and the Infrared Processing and
Analysis Center, funded by the National Aeronautics and Space
Administration and the National Science Foundation. The 2MASS data are
served by the NASA/IPAC Infrared Science Archive, which is operated by
the Jet Propulsion Laboratory, California Institute of Technology,
under contract with the National Aeronautics and Space Administration.
This publication makes use of data products from the Wide-field
Infrared Survey Explorer, which is a joint project of the University
of California, Los Angeles, and the Jet Propulsion
Laboratory/California Institute of Technology, funded by the National
Aeronautics and Space Administration.
\end{acknowledgements}

\facility{K2} \facility{Exoplanet Archive} \facility{IRSA}
\facility{2mass}

\clearpage

\appendix

\section{Mass Estimates}

In order to make the best comparison of rotation properties between
the three age bins, it is desirable to select samples of M dwarfs in
each cluster that span the same mass range.  This is relatively
straight-forward for the Pleiades, Hyades and Praesepe, where all the
stars of interest are on or close to the ZAMS and where theoretical
isochrones have been well-calibrated from comparison to empirical
data.  However, for Upper Sco, the theoretical models are known to
have issues and there are significant disagreements between  models. 
Related to these issues with the models, the age of Upper Sco is a
subject of significant debate and that directly impacts any stellar
mass estimates.  It is also true that for the mass range of interest
to this study, few -- if any -- of the theoretical models predict
accurate $(V-K_{\rm s})_0$ colors for M dwarfs.   This is a problem
for us because we have used $(V-K_{\rm s})_0$ as our primary mass
surrogate in all our papers on the K2 open cluster data.  Despite
these issues, we have attempted to derive reliable mass estimates for
the M dwarfs in the four clusters; we describe our methodology here.

It is first necessary to adopt fundamental parameters for the four
clusters defining the cluster age, distance, metallicity and
reddening.  These parameters are provided in
Table~\ref{tab:basicdata}.   The values in Table~\ref{tab:basicdata}
are those we used in our previous papers on the K2 rotation data for
these clusters (Rebull \etal\ 2016a, 2017, 2018); references for the
adopted ages, distances,  and reddenings can be found in those
papers.   We adopt solar metallicities for Upper Sco and Pleiades
since that is consistent with the overall literature for those
clusters.  Praesepe and the Hyades are generally considered to be
somewhat metal rich (Boesgaard \& Budge 1988; An \etal\ 2007;  Pace
\etal\ 2008; Cummings \etal\ 2017) -- the [Fe/H] = +0.15 value we
adopt is a compromise between the values reported in those papers.  

For the Pleiades, Hyades and Praesepe, we will use theoretical model
isochrones to derive mass estimates based on observables.  Based in
part on arguments presented in David \etal\ (2018) and in part on
evidence provided  below, we adopt the PARSEC models (Chen \etal\
2014) for this purpose. We also will assume that Praesepe and the
Hyades have similar enough properties that we can derive an $M_J$ to
mass transformation for Praesepe, and use that same relation for the
Hyades.

\floattable
\begin{deluxetable*}{lcccc}
\tabletypesize{\footnotesize}
\tablecolumns{5}
\tablewidth{0pt}
\tablecaption{Properties of the Clusters\label{tab:basicdata}}
\tablehead{
\colhead{Cluster Name} &
\colhead{Age } &
\colhead{Distance} &
\colhead{[Fe/H]} &
\colhead{A$_V$}  \\
 & \colhead{(Myr)}& \colhead{(pc)}
& & \colhead{(mag)} }
\startdata
Upper Sco  & 8 Myr  & 140 & 0.0 &  0.7 \\
Pleiades & 125 Myr  & 136 & 0.0 &  0.12 \\
Praesepe & 700 Myr  & 184 & 0.15 &  0.00 \\
Hyades & 700 Myr  &  46 & 0.15 &  0.00 \\
\enddata
\end{deluxetable*}
\noindent

\begin{figure}[ht]
\epsscale{0.9}
\plotone{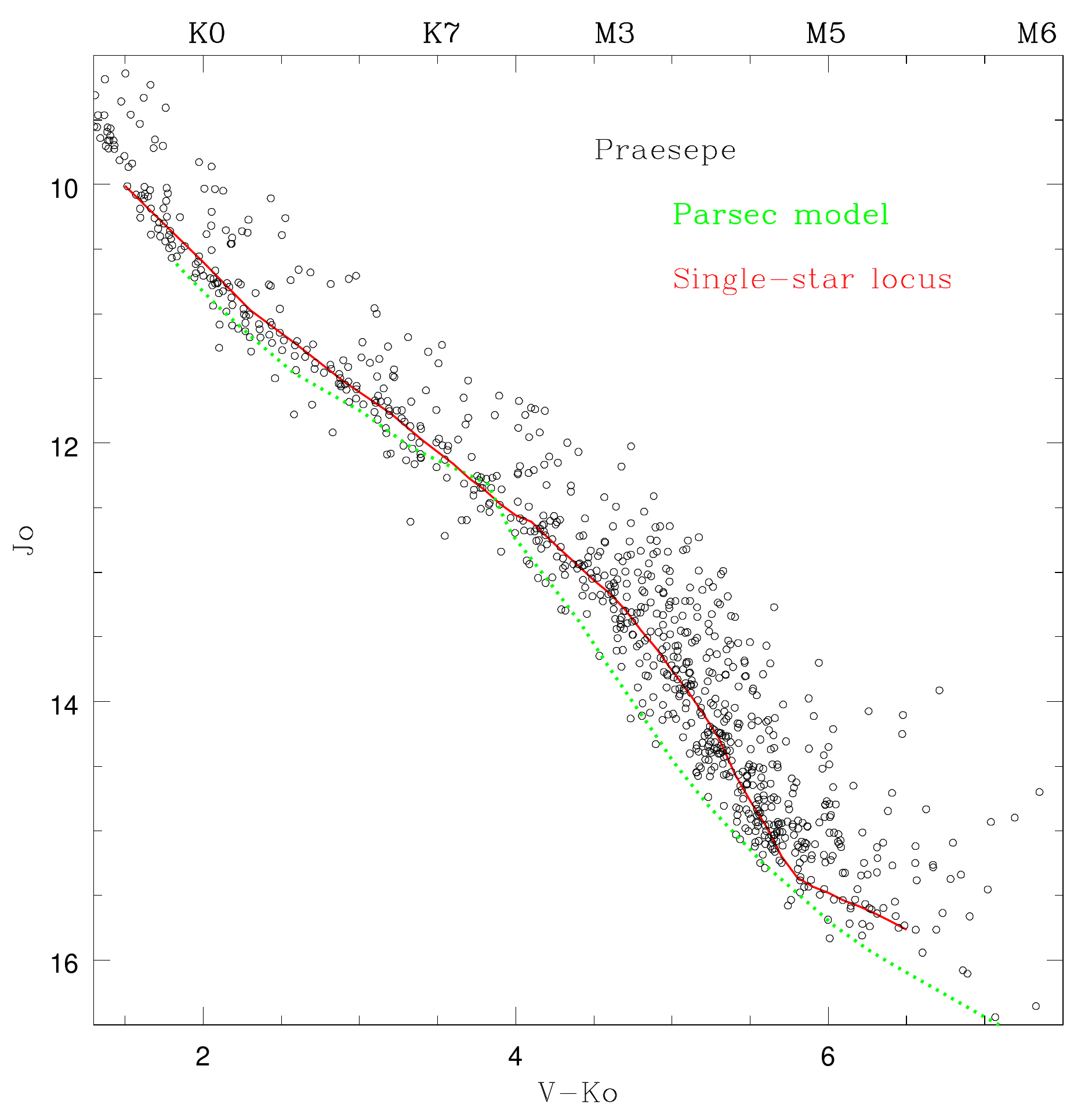}
\caption{CMD for low mass stars in Praesepe for
which we have K2 light curves.  The red, solid curve is our empirical
fit to the single-star locus.  The green, dashed curve is the PARSEC
700 Myr isochrone, assuming a distance of 184 pc, no reddening and
[Fe/H] = +0.1 for Praesepe.
\label{fig:Figure12}}
\end{figure}

For Praesepe, we derive mass estimates in the following manner.  
First, we create a $J$ vs.\ $(V-K_{\rm s})_0$ color magnitude diagram,
as shown in  Figure \ref{fig:Figure12}.   The CMD has a well-defined
single-star locus and a well-populated binary/triple sequence (stars
that fall above the single-star locus, by amounts up to about 1.0 mag
-- which would correspond to a triple system of nearly equal mass
stars).   We fit a curve to the single star locus, as illustrated by
the red line in Figure \ref{fig:Figure12}.  We convert $J$ magnitudes 
to $M_J$ values using the distance and reddening in
Table~\ref{tab:basicdata}.  For every star in the cluster, we assign
it the $M_J$ from this curve corresponding to its $(V-K_{\rm s})_0$.  
For binaries, this means that the mass we associate with the system
corresponds approximately to the mass of the primary star.    Figure
\ref{fig:Figure12} also includes a curve showing the location of the
PARSEC 700 Myr isochrone in the $J$ vs.\ $(V-K_{\rm s})_0$ plane
compared to the actual Praesepe photometry.  While the overall match
is fairly good, there is a significant discrepancy between the
isochrone shape and the locus of Praesepe points amongst the M
dwarfs.  That discrepancy emphasizes why we do not simply use
$(V-K_{\rm s})_0$ color as the link between the theoretical  isochrone
and mass.

\begin{figure}[ht]
\epsscale{0.9}
\plotone{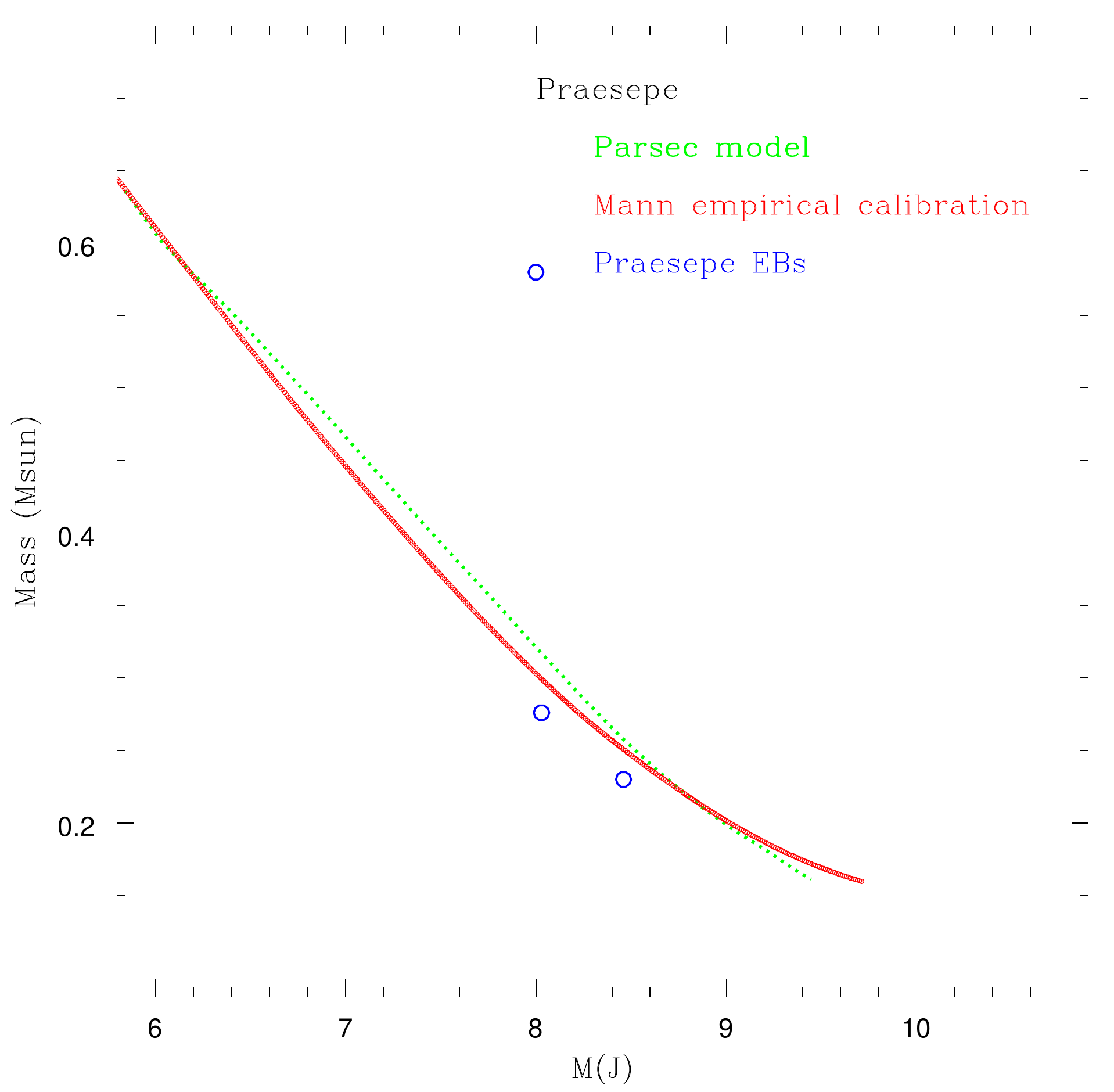}
\caption{Comparison between the predicted mass-$M_J$ correlation for
low mass stars based on the 700 Myr PARSEC isochrone and an empirical
mass-$M_J$ correlation based on the results of Mann \etal\ (2015).
\label{fig:Figure13}}
\end{figure}

As an empirical check on the masses we derive in this way, in Figure
\ref{fig:Figure13} we compare the masses inferred from the PARSEC
model isochrone absolute $J$ magnitude to masses inferred from the
semi-empirical mass-$M_K$ relation derived in Mann \etal\ (2015) from
field stars.   We convert the Mann relation to  mass-$M_J$ using the
observed $J-K_{\rm s}$ colors of the Praesepe stars. Praesepe is old
enough that even the lowest mass M dwarfs with K2 photometry are
essentially on the main sequence, so comparison to a field star
relation should be valid.  Figure 13 shows that the masses derived
from the PARSEC isochrone absolute $J$ magnitude are in good agreement
with the Mann \etal\ relation, with maximum deviations of only a few
hundredths of a solar mass.  Finally, we also show mass/$M_J$  points
for two low mass Praesepe eclipsing binaries identified from K2 data
(Gillen \etal\ 2017).   One of these systems, AD2615, consists of two
nearly equal mass components -- we plot the data point at the average
mass and with $M_{J, \rm primary} = M_{J, \rm system}$ + 0.75.   The
other system, AD3116, consists of a 0.276 \msun\ primary and a brown
dwarf secondary.  We adopt $M_{J, \rm primary} =  M_{J, \rm system}$,
and plot the point at the mass of the primary star.    We do not plot
data for a 3rd Praesepe EB, AD 3814 (Kraus \etal\ 2017), because the 
two stars have masses of 0.38 and 0.21 \msun, and there is no certain
way to estimate the $J$ magnitude of each component.  

\begin{figure}[ht]
\epsscale{0.9}
\plotone{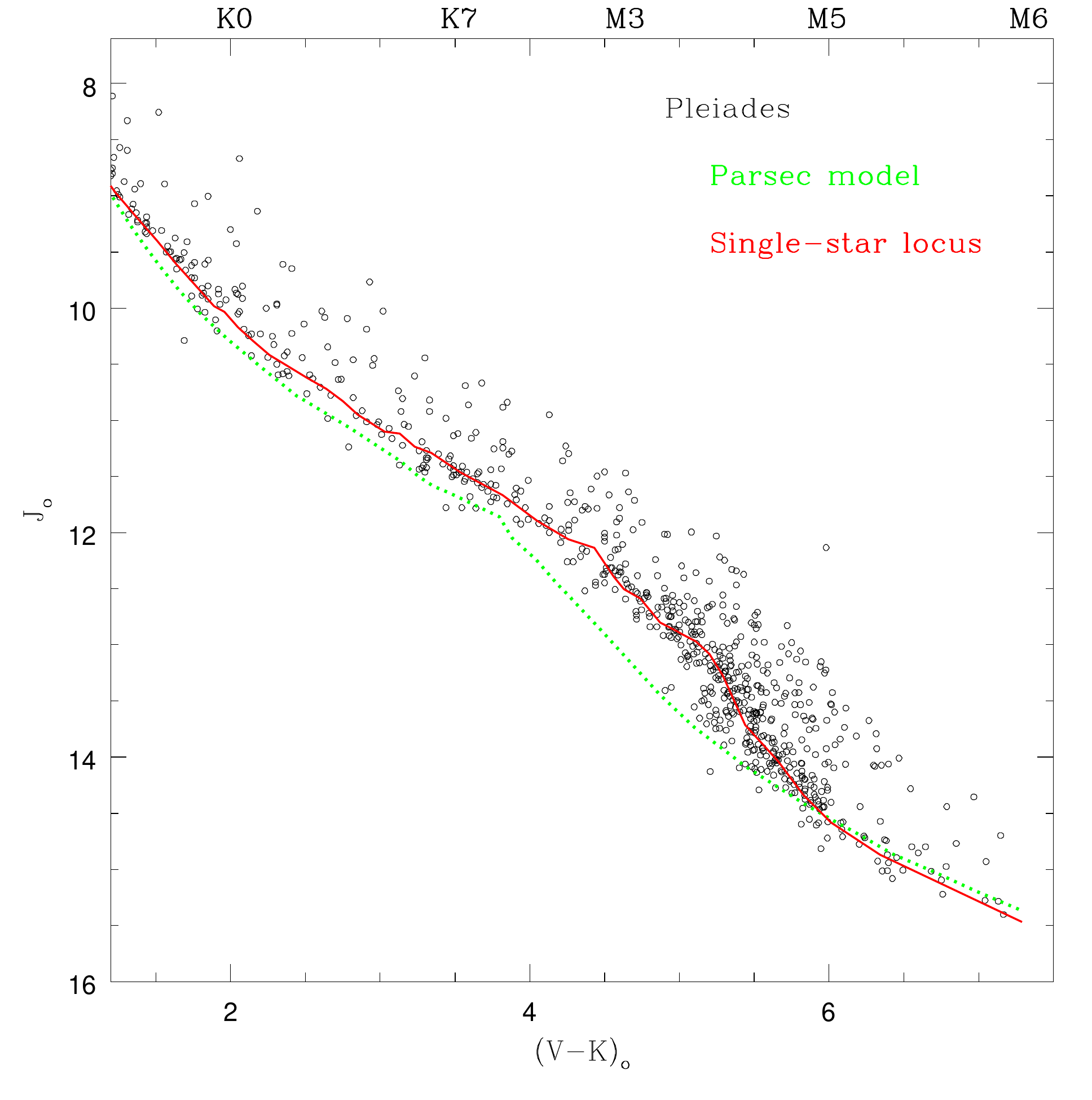}
\caption{CMD for low mass stars in the Pleiades for which we have K2
light curves.  The red, solid curve is our empirical fit to the
single-star locus.  The green, dashed curve is the PARSEC 125 Myr
isochrone, assuming a distance of 136 pc, $A_V$ = 0.12,  and solar
metallicity for the Pleiades.
\label{fig:Figure14}}
\end{figure}

\begin{figure}[ht]
\epsscale{0.9}
\plotone{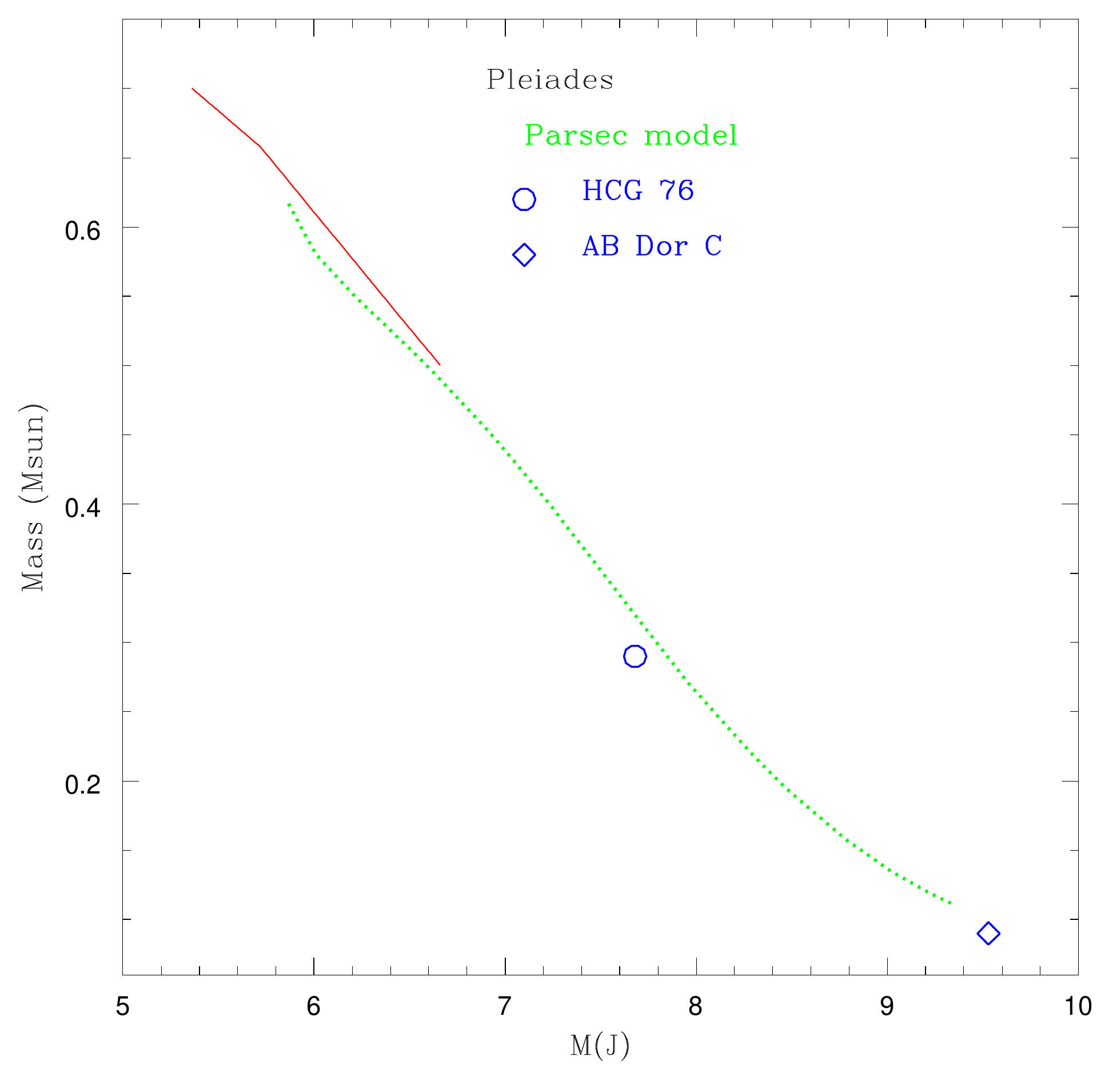}
\caption{Comparison between the predicted mass-$M_J$ correlation for
low mass stars based on the 125 Myr PARSEC isochrone and an empirical
mass-$M_J$ correlation based on the results of Mann \etal\ (2015). 
Also shown are direct mass estimates for the Pleiades dM member HCG 
76 and for the Pleiades age binary star AB Dor C.
\label{fig:Figure15}}
\end{figure}

We followed essentially the same procedure for assigning masses to the
low mass stars in the Pleiades.  Figure \ref{fig:Figure14} shows the
$J$ versus $(V-K_{\rm s})_0$  CMD for the low mass stars observed by
K2.  As for Praesepe, the PARSEC isochrone at Pleiades age provides a
reasonably good fit to the observed photometry for the K dwarfs and
the late M dwarfs, but for 4 $< (V-K_{\rm s})_0 <$ 5.5, the isochrone
is considerably too blue.   Figure \ref{fig:Figure15} shows the mass
versus $M_J$ plot for the Pleiades derived from the PARSEC 125 Myr
isochrone.   As empirical checks on this relation, we also plot as a
solid red line the Mann \etal\ (2015) main sequence relation for just
the mass range where we expect the Pleiades stars to be on or very
near the main sequence.   We also plot the one Pleiades dM EB with
accurate mass estimates where we can also estimate an appropriate
$M_J$,  HCG 76 (David \etal 2016).   We additionally plot data for the
Pleiades age binary star AB Dor C (Close \etal\ 2007; Azulay \etal\
2017; Luhman, Stauffer, and Mamajek 2005).  These empirical checks
confirm that our $M_J$ to mass  conversion derived from the PARSEC
isochrone should provide reasonably accurate masses for the low mass
stars in the Pleiades.

Because theoretical isochrones at young ages are considerably more
uncertain and because of the (related) uncertainty in ages of star
forming regions, we adopted a different procedure for assigning masses
in Upper Sco.   Instead of using the isochrone to define a conversion
from $M_J$ to mass and then using masses derived from binary systems as
an empirical check, for Upper Sco we use the masses from binary
systems to provide the basis for our $M_J$ to mass conversion and use a
theoretical isochrone only as a convenient interpolation scheme
between the  binary star data points.  This is fortunately possible
because there are three Upper Sco dM eclipsing binary systems with
accurate masses and where the components of the binaries are nearly
equal mass (David \etal\ 2016, 2018; Kraus \etal\ 2015). The data for
these three systems are provided in Table~\ref{tab:USCO_EBdata}

\floattable
\begin{deluxetable*}{lcccccc}
\tabletypesize{\footnotesize}
\tablecolumns{7}
\tablewidth{0pt}
\tablecaption{Properties of the Upper Sco dM Eclipsing Binaries\label{tab:USCO_EBdata}}
\tablehead{
\colhead{EPIC Number} &
\colhead{Name } &
\colhead{M$_1$} &
\colhead{M$_2$} &
\colhead{J(system)} & 
\colhead{Spectral Type} &
\colhead{References}  \\
 & & \colhead{(Msun)}& \colhead{Msun}
&  \colhead{(mag)} &  &}
\startdata
204376071 & USco 48 & 0.737 & 0.709 & 9.824 & M1  & David \etal\ (2018) \\
205030103 & UScoCTIO 5 & 0.332 & 0.319 & 11.172 & M4.5 & Kraus \etal\ (2015); David \etal\ (2016) \\
203710387 & \nodata & 0.116 & 0.106 & 12.932 & M5 & David \etal\ (2016) \\
\enddata
\end{deluxetable*}
\noindent

\begin{figure}[ht]
\epsscale{0.9}
\plotone{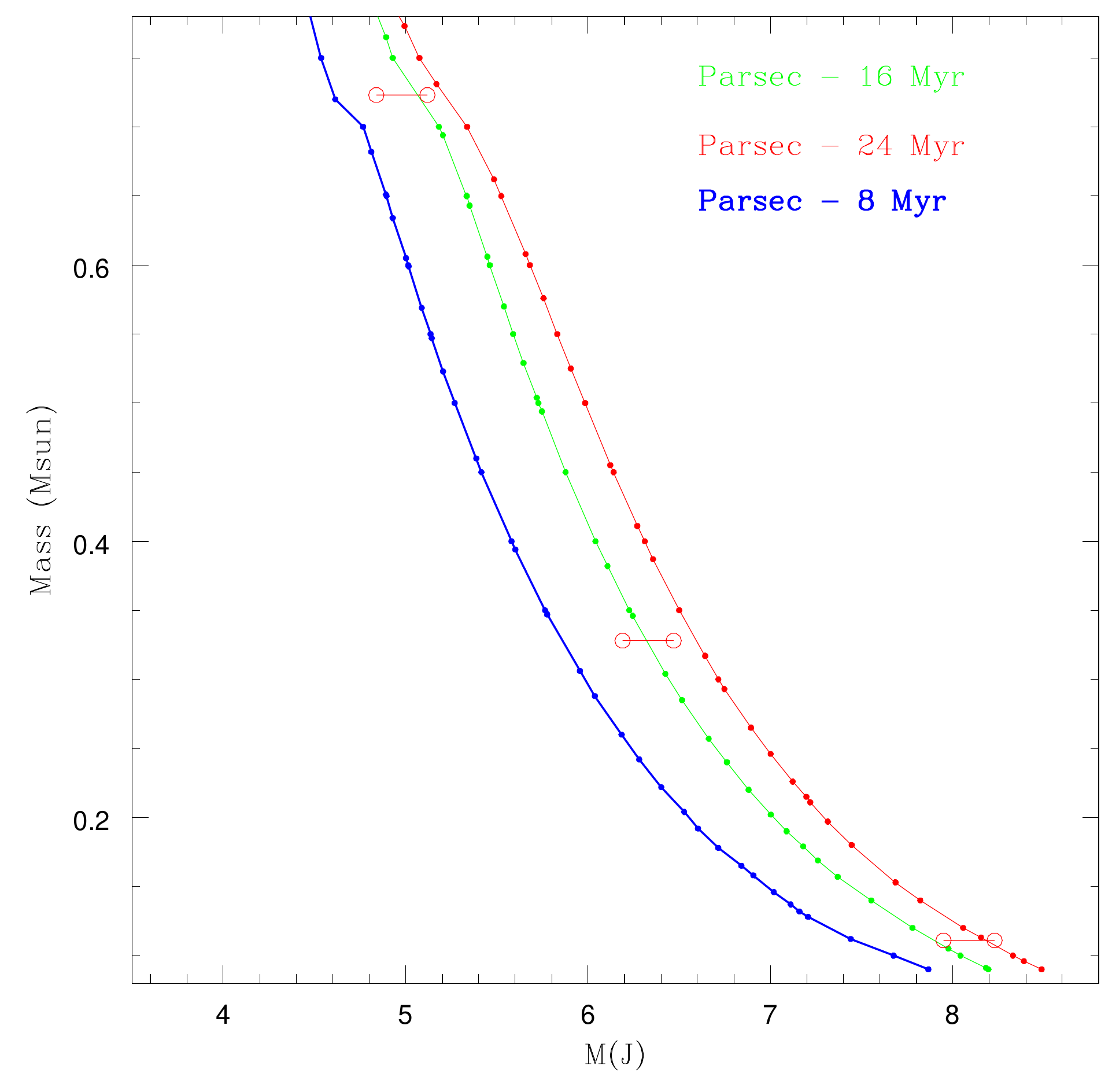}
\caption{Comparison between the predicted mass-$M_J$ correlation for low mass
stars based on the 8, 16 and 24 Myr PARSEC isochrones and data for three
Upper Sco dM eclipsing binary stars with accurate masses (David \etal\ 2016, 2018;
Kraus \etal\ 2015).   The EBs are plotted at both $A_V$ = 0.0 and $A_V$ = 1.0.
\label{fig:Figure16}}
\end{figure}

Figure \ref{fig:Figure16} plots the three Upper Sco EBs in the mass
versus $M_J$ plane, and compares their locations to PARSEC isochrones
for 8, 16 and 24 Myr.   We plot the Upper Sco EBs at both $A_V$ = 0.0
and $A_V$ = 1.0; their actual $A_V$ should lie somewhere  within that
range.  For each star, we use the average of the two masses, and we
add 0.75 mag to the observed $J$ magnitude to get a magnitude
corresponding to just one of the stars.  It can be seen that the
PARSEC isochrones have a similar shape in this plane to the three
fiducial points, with the 16 Myr isochrone most nearly matching the
three fiducials.   Because it is thought that magnetic fields and
starspots (not included in the PARSEC models) may significantly
influence the observed isochrone magnitudes and colors, we are not
advocating that this is favors a 16 Myr age for Upper Sco.   We simply
advocate that the 16 Myr isochrone provides a convenient means to
place our K2 Upper Sco stars onto a mass scale that is consistent with
the empirical data provided by the EBs.

Tables~\ref{tab:USCO_data}, \ref{tab:PLE_data}, and \ref{tab:PrHya_data}
provide the colors, periods and mass estimates for the $M <$ 0.32 \msun\ stars
that are analysed in the main text of the paper.  The colors and periods are
the same as originally reported in Rebull \etal\ (2016a, 2017 and 2018a).

\floattable
\begin{deluxetable*}{lccccc}
\tabletypesize{\footnotesize}
\tablecolumns{6}
\tablewidth{0pt}
\tablecaption{Upper Sco M $<$ 0.32 \msun\ Stars with K2 Periods\label{tab:USCO_data}}
\tablehead{
\colhead{EPIC Number} &
\colhead{(V-K)$_o$ } &
\colhead{P$_1$} &
\colhead{P$_2$} &
\colhead{Mass} & 
\colhead{Disk?} \\ 
 & & \colhead{(days)}& \colhead{days}
&  \colhead{(\msun)} &  }
\startdata
202515599 & 6.460 & 0.5851 & \nodata & 0.123 & \\
202632400 & 6.073 & 1.3723 & \nodata & 0.182 & \\
202638454 & 6.492 & 0.2487 & 0.4455 & 0.119 & \\
202709862 & 5.655 & 0.9952 & 2.0106 & 0.289 & \\
202793212 & 5.590 & 2.1634 & \nodata & 0.310 & \\
202795175 & 5.883 & 0.8361 & \nodata & 0.224 & \\
202828127 & 6.388 & 1.1663 & \nodata & 0.133 & Y \\
202873945 & 6.161 & 0.6258 & \nodata & 0.166 & \\
202879519 & 6.011 & 5.0729 & \nodata & 0.193 & \\
\enddata
\tablenotetext{a}{This table is available in its entirety in the online
version. A portion is shown here to demonstrate its form and content.}
\end{deluxetable*}

\floattable
\begin{deluxetable*}{lcccc}
\tabletypesize{\footnotesize}
\tablecolumns{5}
\tablewidth{0pt}
\tablecaption{Pleiades M $<$ 0.32 \msun\ Stars with K2 Periods\label{tab:PLE_data}}
\tablehead{
\colhead{EPIC Number} &
\colhead{(V-K)$_o$ } &
\colhead{P$_1$} &
\colhead{P$_2$} &
\colhead{Mass} \\
 & & \colhead{(days)}& \colhead{days}
&  \colhead{(\msun)}   }
\startdata
210769047 & 5.705 & 0.6439 & \nodata & 0.194 \\
210791550 & 5.471 & 0.3727 & \nodata & 0.248 \\
210815768 & 5.867 & 0.2999 & \nodata & 0.163 \\
210822528 & 5.818 & 0.9149 & \nodata & 0.171 \\
210832378 & 5.462 & 0.4656 & \nodata & 0.251 \\
210833622 & 5.452 & 1.0465 & \nodata & 0.254 \\
210854098 & 6.079 & 0.3451 & \nodata & 0.137 \\
210860152 & 5.486 & 0.6299 & \nodata & 0.243 \\
210866482 & 5.482 & 0.3228 & \nodata & 0.245 \\
210871940 & 5.313 & 0.2773 & \nodata & 0.316 \\
\enddata
\tablenotetext{a}{This table is available in its entirety in the online
version. A portion is shown here to demonstrate its form and content.}
\end{deluxetable*}

\floattable
\begin{deluxetable*}{lccccc}
\tabletypesize{\footnotesize}
\tablecolumns{6}
\tablewidth{0pt}
\tablecaption{Praesepe/Hyades M $<$ 0.32 \msun\ Stars with K2 Periods\label{tab:PrHya_data}}
\tablehead{
\colhead{EPIC Number} &
\colhead{(V-K)$_o$ } &
\colhead{P$_1$} &
\colhead{P$_2$} &
\colhead{Mass} & 
\colhead{Cluster} \\ 
 & & \colhead{(days)}& \colhead{days}
&  \colhead{(\msun)} &  }
\startdata
210468157 & 5.515 & 1.3950 & \nodata & 0.256 & H \\
210540496 & 6.162 & 0.3409 & \nodata & 0.178 & H \\
210640966 & 5.554 & 2.7156 & 2.5569 & 0.247 & H \\
210674207 & 5.469 & 0.9868 & 1.0488 & 0.268 & H \\
210700098 & 5.361 & 2.3441 & \nodata & 0.302 & H \\
210742017 & 5.690 & 2.8806 & \nodata & 0.213 & H \\
210742592 & 5.341 & 0.8036 & \nodata & 0.310 & H \\
210744677 & 5.420 & 0.4813 & 0.5136 & 0.281 & H \\
210754620 & 5.381 & 0.6339 & \nodata & 0.294 & H \\
210769813 & 5.340 & 1.8632 & \nodata & 0.310 & H \\
\enddata
\tablenotetext{a}{This table is available in its entirety in the online
version. A portion is shown here to demonstrate its form and content.}
\end{deluxetable*}

\section{Reality of the Spike in Upper Sco M Dwarf Rotation Periods
Near $P$=1.8 days}

Close examination of Figures \ref{fig:Figure4} and \ref{fig:Figure5} 
of \S 3 seems to show a peculiar linear, approximately horizontal
structure for $5 < (V-K_{\rm s}< 6.5$ and periods near but slightly
less than 2 days (log $P$=0.3).  We worried that this might be a sign
that some artifact in the K2 data acquisition or pipeline processing
was imposing this period on the data and therefore at least some of
these periods are not real.   We were particularly concerned about an
artifact at or near $P$=2 days, because that is the period at which K2
dumps angular momenta that have built up in their reaction wheels (van
Cleve \etal\ 2015).   Small pointing jumps are thus likely to occur
with that periodicity, which could be imprinted on the  time series
photometry.

\begin{figure}[ht]
\epsscale{1.0}
\plotone{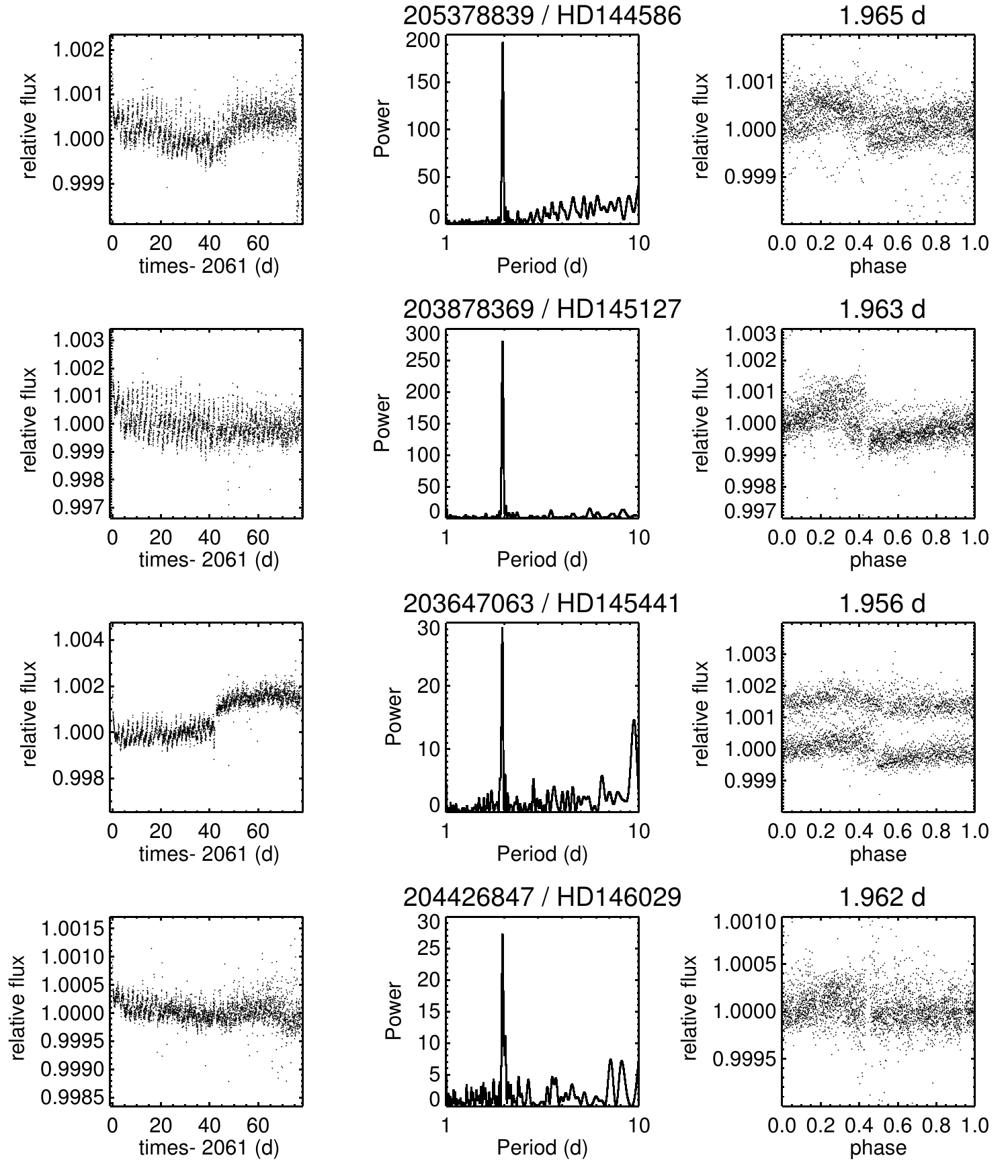}
\caption{Four examples of light curves with spurious $\sim$2d periods.
First column: light curve; second column: power spectrum; third
column: phased light curve for the $\sim$2d period peak in the
periodogram. Note the near discontinuity in the phased light curves,
often coinciding with a small gap in the data. We discarded LCs with
this kind of behavior as likely instrumental.
Specific stars are: EPIC 205378839/HD144586,
EPIC 203878369/HD145127,
EPIC 203647063/HD145441,
EPIC 204426847/HD146029.
\label{fig:Figure17}}
\end{figure}

We had noticed a concentration of periods near $P$=2.0 days while
preparing the data tables for Rebull \etal\ (2018a).   In fact, our
original set of periods had even more stars near this period, making
plots of period versus color appear even more anomalous. Therefore, in
the context of Rebull \etal\ (2018a), we reviewed extensively the
light curves between 1.5 and 2.5 days. We think it likely that there
is indeed an instrumental or processing artifact near 2.0 days; see
Fig.~\ref{fig:Figure17} for four examples of light curves with
apparent 2 day periods that we discarded as likely spurious. (these
periods were not reported in Rebull \etal\ 2018a).  Note that these
phased light curves are almost discontinuous.  They are also extremely
low amplitude.   We, in fact, believe that the stars where one sees
this 1.96 day instrumental period are almost exclusively stars that
are intrinsically non-variable.   The imposed 2 day period only
appears with large enough signal to be detected in the Lomb-Scargle
periodogram for such non-variable stars.   Examination of the K2 light
curve data files shows that there are indeed small position shifts at
the times when momentum dumps are flagged, and that these do occur at
a 1.96 day cadence.

\begin{figure}[ht]
\epsscale{1.0}
\plotone{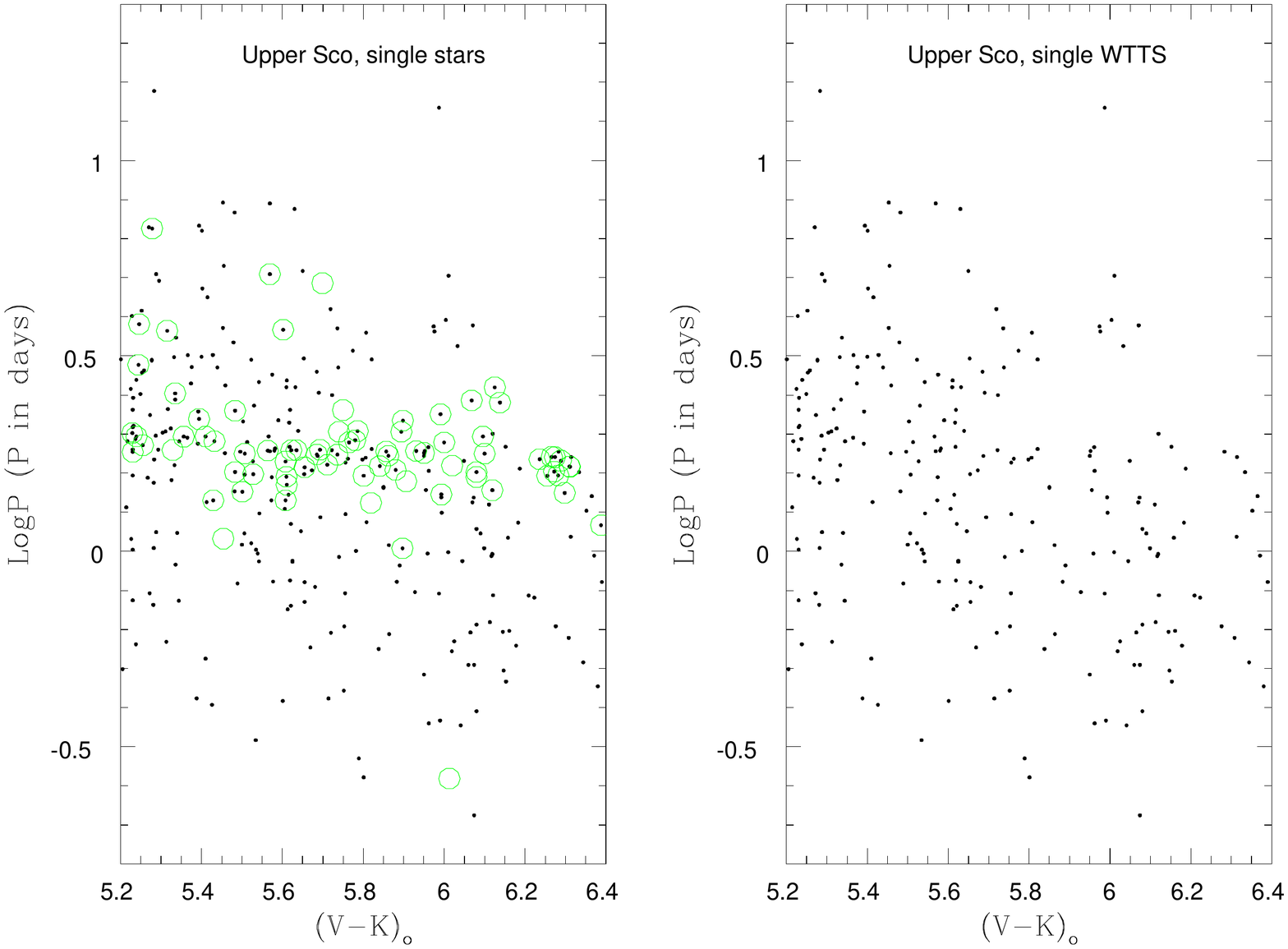}
\caption{(LHS) Periods for Upper Sco M $<$\ 0.32 \msun\ single stars,
with black dots corresponding to stars without IR excesses and green
circles corresponding to stars with IR excesses.  The periods for the IR-excess
stars (CTTS)
concentrate around a period a little less than two days, whereas
the stars without disks have a much more scattered period distribution.
(RHS) Same plot, but now plotting only the stars without IR excesses.
There is no obvious concentration of points near logP = 0.3.
\label{fig:Figure18}}
\end{figure}

Figures \ref{fig:Figure5} and \ref{fig:Figure6}  of \S 3 were
constructed after our removal of all the stars we believed had
instrumentally-imposed 2 day periods.   We believe that all of the
periods we retained in Rebull \etal\ (2018a) and that we have used in
this paper are physical.  One reason for this belief is that most or
all of the remaining excess in periods near 2.0 days is attributable
to CTTS; any instrumentally-induced periodicity should have no
knowledge of whether the star is a CTT or a WTT.   We illustrate this
point in Fig.~\ref{fig:Figure18}a, which replots the Upper Sco
rotation data for only stars with single K2 periods, and where we
highlight (as green circles) stars with certain IR excesses (i.e.,
they are CTTS).  The CTTS are concentrated around $P \sim$1.8 days
(log$P$=0.25). If one removes the CTTS from the diagram, as shown in 
Fig.~\ref{fig:Figure18}b, there is no longer any significant hint of
an excess.

\begin{figure}[ht]
\epsscale{1.0}
\plotone{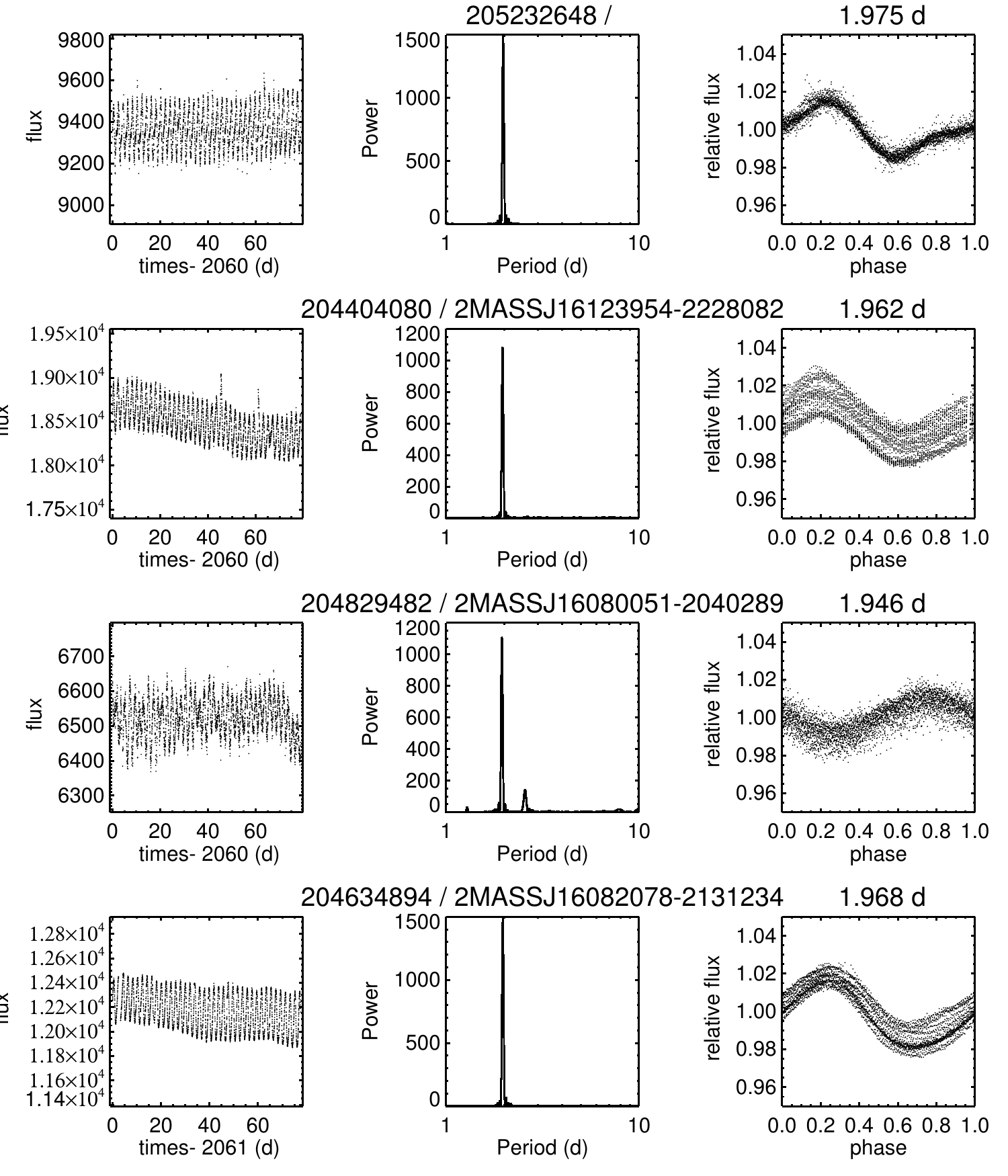}
\caption{Four examples of light curves (from disk-free stars) with
intrinsic $\sim$2d periods. First column: light curve; second column:
power spectrum; third column: phased light curve for the $\sim$2d
period peak in the periodogram.  We retained LCs with this kind of
behavior as likely real.
Specific stars are: EPIC 205232648,
EPIC 204404080/2MASSJ16123954-2228082,
EPIC 204829482/2MASSJ16080051-2040289,
EPIC 204634894/2MASSJ16082078-2131234.
\label{fig:Figure19}}
\end{figure}

\begin{figure}[ht]
\epsscale{1.0}
\plotone{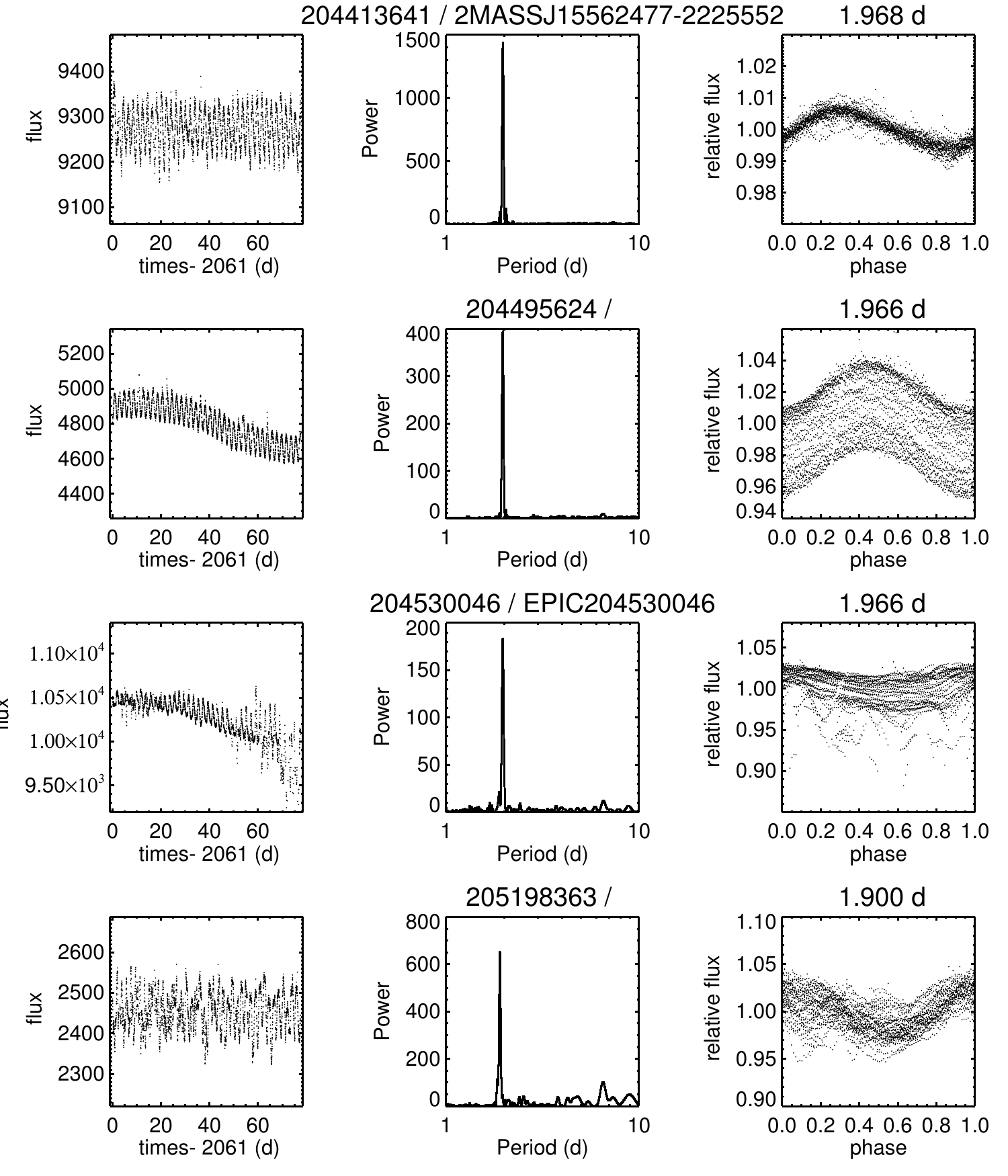}
\caption{Four examples of light curves (from disked stars) with
intrinsic $\sim$2d periods. First column: light curve; second column:
power spectrum; third column: phased light curve for the $\sim$2d
period peak in the periodogram.  We retained LCs with this kind of
behavior as likely real.
Specific stars are: EPIC 204413641/2MASSJ15562477-2225552 (note that this star also
appears in campaign 15, with the same period),
EPIC 204495624,
EPIC 204530046 (a dipper), and
EPIC 205198363.
\label{fig:Figure20}}
\end{figure}

In contrast to the stars with instrumentally-imposed periods,
Fig.~\ref{fig:Figure19} and \ref{fig:Figure20} show that stars that 
have intrinsic periods near $\sim$2 d have normal light curves for
young, low mass stars. Fig.~\ref{fig:Figure19} are stars without disks
and Fig.~\ref{fig:Figure20} are stars with disks. Note that the phased
light curves are completely different than the ones in
Fig.~\ref{fig:Figure17}, even for those stars with disks; the phased
light curves are continuous, often sinusoidal. 

\begin{figure}[ht]
\epsscale{1.0}
\plotone{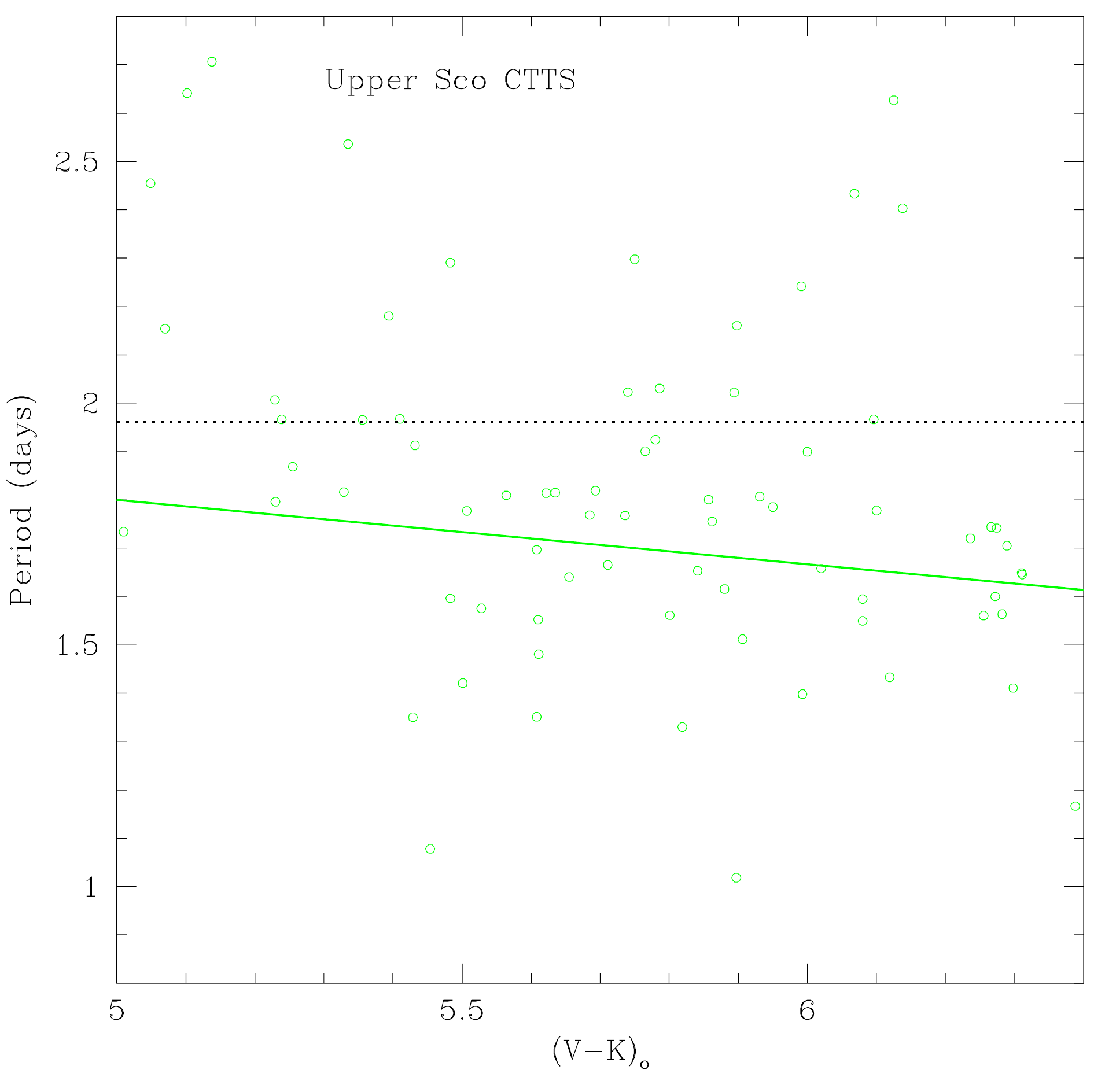}
\caption{Plot of the periods for the Sco CTTS with M $<$ 0.32 \msun\ vs. V-K.  The
solid green lines is the linear
fit to these data; the dashed black line corresponds to the 1.96 day cadence for
the standard K2 angular momentum dumps.   The typical accuracy with which
the periods are determined is similar or smaller than the size of the
data points.  See text for discussion.
\label{fig:Figure21}}
\end{figure}

Finally, we have fit a line to the periods for just the disked stars
(using outlier-resistant linear fitting robust\_linefit.pro from the
IDL astro library) for 5$<(V-K_{\rm s})_0<$6.4 and 1$<P<$3. The fit,
shown in  Figure~\ref{fig:Figure21}, has small but significantly
non-zero slope ($-$0.0724$\pm$0.0241).  We also provide a dotted line
in the figure at the instrumentally-induced period of 1.96 days. There
are three take-aways from this figure:
\begin{itemize}
\item The CTTS periods do not concentrate around the 1.96 day
instrumentally-induced periiod;
\item While the dM CTTS periods are concentrated to a relatively small
range in period, their disperion in period is MUCH larger than the
uncertainty in the individual periods;
\item The CTTS periods are a (weak) function of color, with longer
mean periods at higher mass.
\end{itemize}

All three of those points argue against the apparent linear
arrangement of periods in Figures \ref{fig:Figure4} and
\ref{fig:Figure5} of \S 3 being an artifact, and instead argue that
the preference for periods near 1.8 days for low mass CTTS at Upper
Sco age must be due to a physical process such as the disk-locking
mechanism of Ghosh \& Lamb (1979).

\section{How Complete is Our Identification of dM Binaries with K2}

Two items of information provided in Sec. 3.1 seem to be in conflict. 
The first is our statement that we detect periods for about $\sim$90\%
of the low mass members in these young star regions.   The second item
derives from examination of the color-magnitude diagrams (CMDs) for
the Pleiades and Praesepe in the lower half of Figure 2.  Those CMDs
were meant to highlight that the K2 dM stars with two periods are,
with only a few exceptions, well displaced above the single-star
locus, therefore confirming that they are also photometric binaries. 
What also  appears to be true based on those two plots, however, is
that an approximately equal number of low mass photometric binaries
(stars displaced more than 0.4 mag above the single star locus) in
these two clusters have only one detected  period.   Why do we not
detect a second period in these stars?   Does this issue significantly
affect any of our conclusions in the main text concerning the rotation
rates of single and K2-detected binary stars?    We address these
issues here.  

There are several potential explanations for why there are so many dM
stars that appear to be photometric binaries but which have only one
period detected with K2:
\begin{itemize}

\item  Non-member contamination;

\item  Our census of the K2 light curve data was incomplete - we failed
   to identify some of the stars with two real periods;

\item  There are single stars masquerading as photometric binaries due
   to bad photometric data;

\item  There are binary members with unusually little photometric variability;

\item  There are secondary stars that are too faint to detect their variability
   with K2,  but still bright enough to cause a a $\Delta$V $>$ 0.4 mag displacement above
   the single star locus in the V vs. \vmk\ CMD.

\end{itemize}

We discuss each of these potential explanations in order.   We only discuss
the Pleiades because it is younger and closer, both of which enhance our ability
to quantitatively test how these mechanisms may affect our ability to
identify a complete set of M dwarf binaries with K2.

\subsection{Non-Member Contamination}

\begin{figure}[ht]
\epsscale{0.9}
\plotone{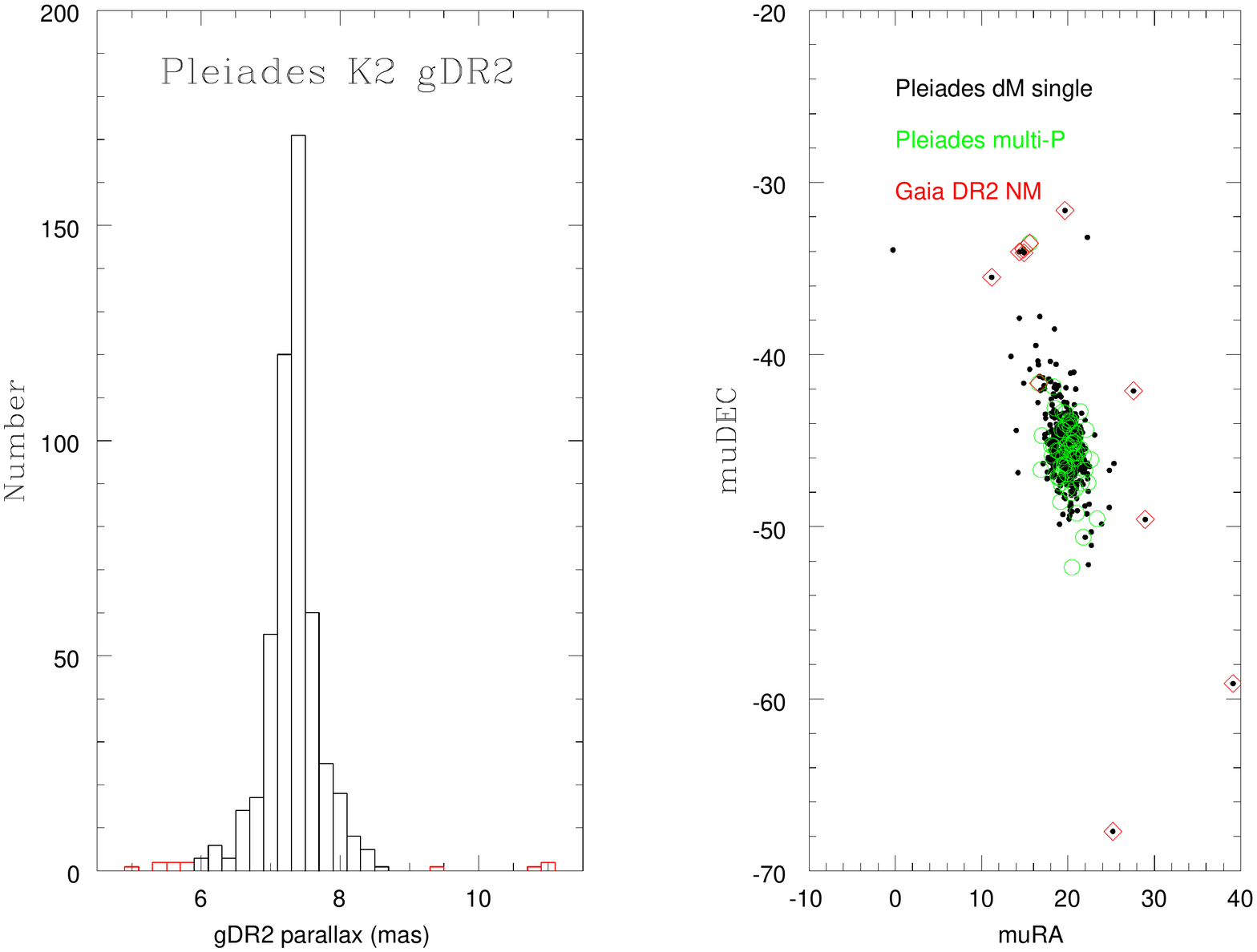}
\caption{(left) Histogram of Gaia DR2 parallaxes for candidate
M $<$ 0.32\msun\ candidate Pleiades members from Rebull \etal\
(2016).  Boxes shown in red are stars we consider probable
non-members based on their DR2 parallaxes.
(right) Vector point diagram for candidate M $<$ 0.32\msun\ 
Pleiades members from Rebull \etal\ (2016) based on Gaia DR2 proper
motions.  K2 single stars are shown as black dots; K2 binary stars are
shown as green open circles; and stars with discrepant DR2 parallaxes
are shown as red diamonds.
\label{fig:FigureC1}}
\end{figure}

A foreground, non-member of the cluster could be located above the
single star locus in the cluster CMD even if it is, in fact, a single
star.  Such a star could be categorized as  a photometric-binary if
plotted in the Pleiades CMD  but as a K2 single assuming it did have a
detectable period.  Fortunately, Gaia DR2 has recently provided us
with accurate parallaxes and proper motions for the great majority of
the Pleiades K2 stars, thus allowing us to better ascertain membership
status. Figure \ref{fig:FigureC1}a shows a histogram of the Gaia DR2
parallaxes for the Pleiades M dwarfs in Figure 2; Figure
\ref{fig:FigureC1}b  provides the corresponding vector point diagram
for those same stars.  The stars with discrepant parallaxes in Figure
C1a are highlighted in Figure C1b by red circles.  In most cases, the
two measures agree - that is, stars with discrepant parallaxes usually
also have discrepant proper motions.  The number of possible
non-members flagged in this way is relatively small, amounting to only
about 2\% of the stars.  More importantly, only three of these stars
falls significantly above the single star locus in Figure 2, one of
which we have identified as a K2 binary and the other two as K2
singles.   Therefore, non-member stars do not significantly
contaminate the dM photometric binary  census in the Pleiades.

\subsection{Previously Missed K2 Binaries}

We were careful to try to identify all stars with more than one real
period in our original K2 Pleiades paper (Rebull \etal\ 2016a). 
However, with more than a thousand stars and half a dozen light curve
versions, there is inevitably room for improvement.  Also, additional
light curve versions have become available, and our own experience in
sifting through the data has evolved with time.  Therefore,  after we
had written the initial draft of this paper and identified the
correlation between binarity and rotation, we reexamined the light
curve data for all of the $M <$0.32 \msun\ K2 stars in the
Pleiades.   We attempted both to identify new K2 binaries and to
determine if any of our previously identified K2 binaries should in
fact have been assigned only one period.  While we did identify a few
stars in both categories, the total number of changes was quite modest
and the new plots and statistical measures (which are given in the
main text) do not differ significantly from our original results.  
The changes to the single and K2 binary categorizations compared to
Rebull \etal\ (2016a) are summarized in Table C1.   While some future
reprocessing of the K2 data may yield new results, we are confident
that with the existing K2 light curves and analysis techniques our
current census of single and binary dMs is as good as we can produce.

\floattable
\begin{deluxetable*}{lcccc}
\tabletypesize{\footnotesize}
\tablecolumns{5}
\tablewidth{0pt}
\tablecaption{Changes to Pleiades Rotation Period Relative to Rebull \etal 2016a\label{tab:Ple_Changes}}
\tablehead{
\colhead{EPIC Number} &
\colhead{(V-K)$_o$ } &
\colhead{P$_1$} &
\colhead{P$_2$} &
\colhead{Comments}  \\
 & & \colhead{(days)}& \colhead{days}
&    }
\startdata
211040347 &  6.11 &  0.3337 & 0.3650   &   added P2 \\
211111473 &  5.90 &  0.2628 & 0.2460   &   added P2  \\
211125179 &  5.59 &  0.3484 & 0.3393   &   added P2  \\
211030074 &  5.89 &  0.1295 & 0.1367   &   added P2  \\
210860152 &  5.49 &  0.6300 & 0.6405   &   added P2  \\
211079163 &  5.68 &  0.2350 & 0.1442   &   added P2  \\
211103322 &  5.56 &  0.3863 & \nodata   &   orig P2 that of companion star \\
211088777 &  5.77 &  1.5972 & \nodata   &   deleted P2  \\
210940129 &  6.07 &  0.7270 & \nodata   &   deleted P2   \\
\enddata
\end{deluxetable*}
\noindent

\subsection{Photometric Uncertainties That Impair Identification of Photometric Binaries}

The \vmk\ colors we have used for the low mass Pleiades and Praesepe
members are in some cases measured and in some cases inferred from
other photometric systems. For every star, the optical and IR
photometry were obtained non-simultaneously, so the photometric
variability of these young, spotted stars may in some cases yield
estimated colors that differ significantly from what would be obtained
with simultaneous data.   Therefore, it is possible that some of the
stars that are apparent  photometric binaries may instead be stars
with unusually large errors in their estimated \vmk\ colors.

The Gaia DR2 data release provides a means to assess this issue
because DR2 provided photometry in three bands -- a very broad G band,
and narrower $G_{BP}$ (blue) and  $G_{RP}$ (red) bands. For each star,
the data in these bands were obtained simultaneously -- and they are
from a space environment which allows significantly better photometric
precision than is true from ground-based surveys.  And, compared to
the heterogenous nature of our \vmk\ colors, these Gaia data are quite
homogeneous.

\begin{figure}[ht]
\epsscale{0.9}
\plotone{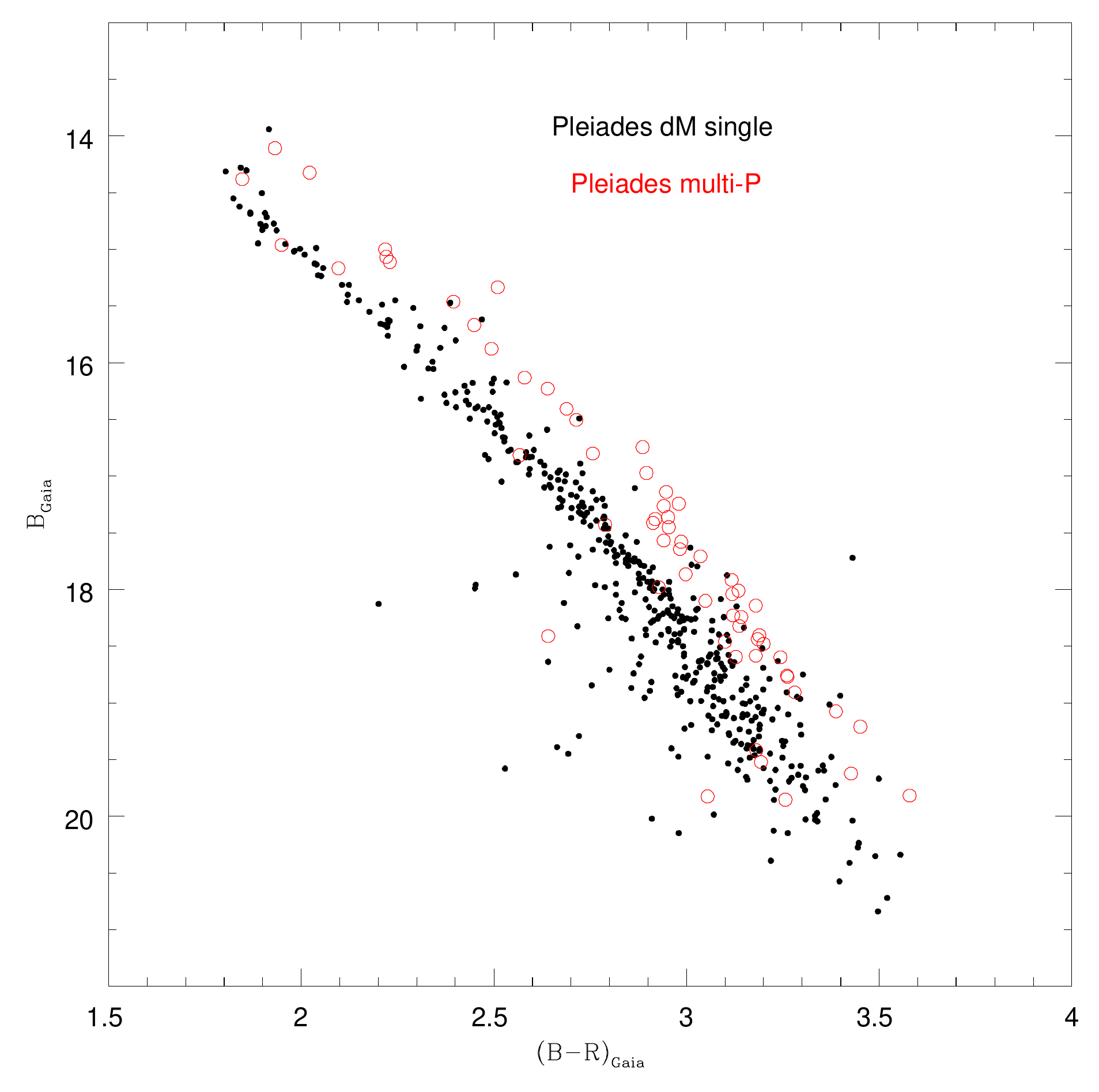}
\caption{Gaia DR2 $G_{BP}$ vs.\ $G_{BP}-G_{RP}$ CMD for the Pleiades
dM stars from Figure 2. Black dots are stars with only a single period
identified in its K2 light curve; red open circles are stars
identified with two periods from the K2 data.
\label{fig:FigureC2}}
\end{figure}

Figure \ref{fig:FigureC2} shows a Gaia DR2 $G_{BP}$ vs.\
$G_{BP}-G_{RP}$ CMD for the Pleiades dM stars from Figure 2; the K2
binary stars are shown as red circles.  For the bright portion of
these stars ($G_{BP}<$ 17.5, corresponding to $V \simlt$ 17.1),
these data suggest that photometric uncertainties in the \vmk\  colors
may indeed have led to the appearance of a larger photometric binary
population than is in fact the case.   That is, a much smaller
fraction of the photometric binaries with $G_{BP} <$ 17.5 in Figure
\ref{fig:FigureC2} are K2 singles as compared to the same set of stars
with $V <$ 17.1 in the lower-left panel of Figure 2.

Fainter than $G_{BP} =$ 17.5, the situation is not so clear.  The
ratio of K2 binaries to all stars near the upper envelope to the
Pleiades locus still seems to be larger than for the same mass range
($V >$ 17.1) in the $V$ vs. \vmk\ diagram, but there is also a large
population of stars scattered well to the blue of the main locus of
Pleiades stars, and the single star locus seems to broaden
considerably for $G_{BP} >$ 17.5.   We believe the increased
scatter in the Pleiades photometry for $G_{BP} >$ 17.5 arises from
a combination of the faintness of these stars and from known current
issues with the DR2 ($G_{BP}-G_{RP}$) colors thought possibly to arise from
problems with sky subtraction in regions where there are bright stars
or complex backgrounds.   Given these issues, it isn't possible to
draw any conclusions for the $M <$ 0.32 \msun\ mass range concerning
the photometric binaries that are K2 singles.   However, based on what
is seen for $G_{BP}$\ $<$ 17.5, we believe it is likely that the
actual number of photometric binaries that are not detected as K2
binaries will be significantly lower once we do have better
photometry.

\subsection{A Population of Photometrically Stable Secondaries?}

If viewed from near the rotational pole, even a very
non-axisymmetrically spotted star will show very little photometric
variability.  This presumably provides a partial explanation for why
we do not detect any rotation period for a small fraction ($<$ 10\%)
of the late-type Pleiades candidate members for which we have K2
data.  If the rotational axes of stars in binaries are not perfectly
aligned, then binaries would have two chances of near polar alignment
to our line of sight.   However, while this might contribute to the
presence of some K2 single stars among the photometric binaries, it
could not explain the very large number shown in the lower-left panel
of Figure 2.

A potentially more serious problem could occur if photometric
variability amplitude and rotation period were strongly linked for
young stars, in the sense of lower amplitudes for slower rotators.  In
that case, the K2 singles that are photometric binaries could be
systems with one component that is unusually slowly  rotating and
hence has little periodic signature in its K2 light curve.  If this
explanation were valid, it could make our finding of faster rotation
in binaries relative to singles be a selection effect, in that our
sample is missing the most slowly rotating binary star members.

\begin{figure}[ht]
\epsscale{0.9}
\plotone{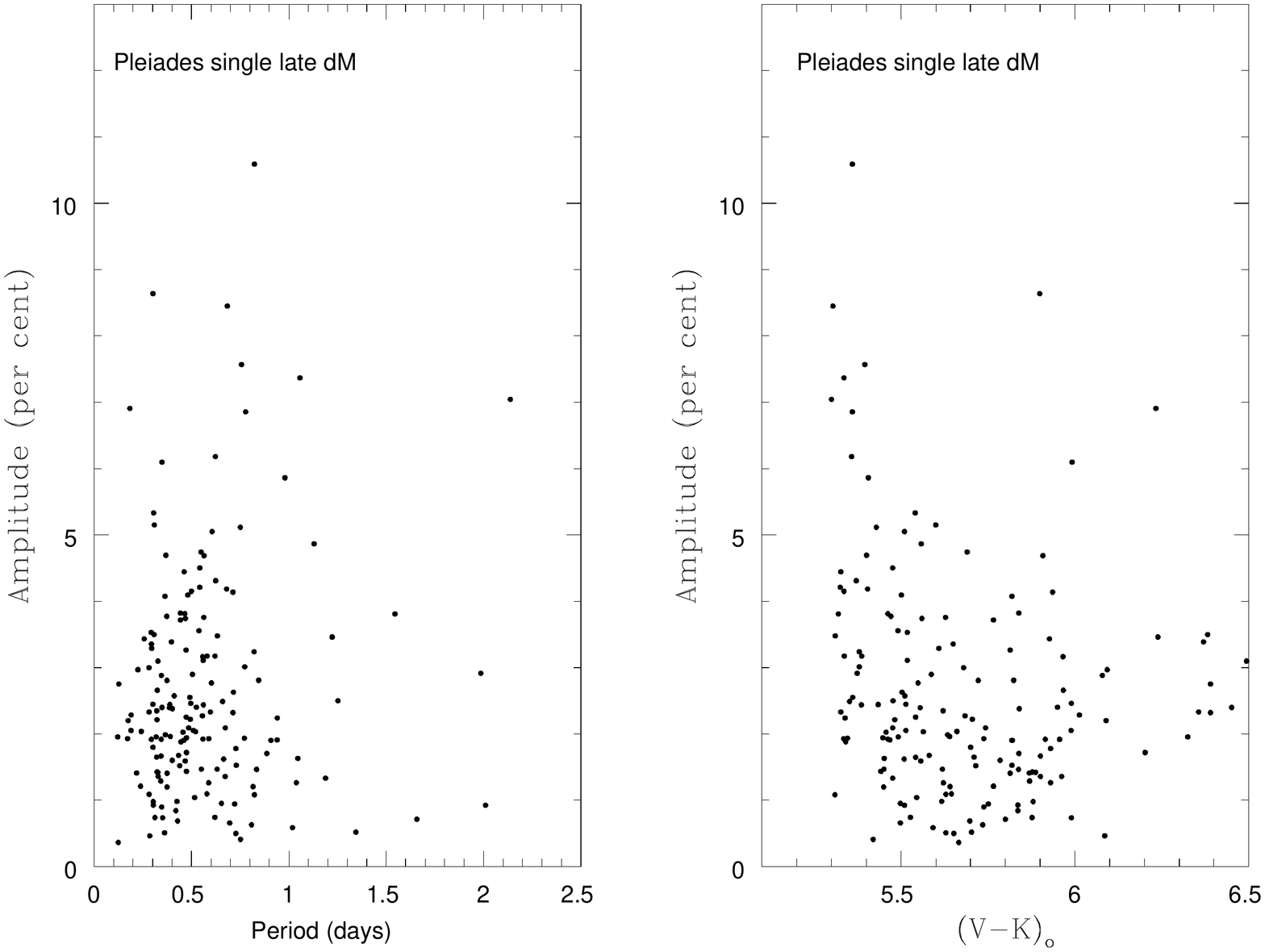}
\caption{(Left) Full amplitude of the phased K2 light curves of the
single, late-type Pleiades dM stars versus our derived periods. 
(Right) The same photometric amplitudes but now plotted versus the
\vmk\ color of the star.  Both plots primarily show a large scatter at
any given period or \vmk\ color, with no significant correlation
between amplitude and period or color.
\label{fig:FigureC3}}
\end{figure}

To test this hypothesis, we have carefully remeasured the photometric
amplitudes of all the $M <$ 0.32 \msun\ Pleiades dM stars for which we
detect only one period.  We plot these amplitudes versus both period
and \vmk\ color in Figure \ref{fig:FigureC3}.  Those plots mostly show
that there is a wide spread in photometric amplitude at a given period
and at a given color.  The range in amplitude by far exceeds any
possible correlation between amplitude and period or amplitude and
color.  One thing that is evident (RHS of Figure C3) is that for the
faintest/reddest stars, we probably are only detecting the largest
amplitude variables -- a likely occurrence since these stars are near
the faint limit for identifying periods with K2.

Figure \ref{fig:FigureC3} thus provides no basis for there being a
hidden population of slowly rotating, low amplitude companions that
our K2 data would have failed to detect.

\subsection{Secondaries Too Faint for K2 to Detect Periods}

K2 can detect rotation periods in remarkably faint stars.  The
faintest (lowest mass) star for which we report a detected period in
the Pleiades has V $\sim$ 20.5,  or K2mag $\sim$17.3.   It is likely
that we only detect periods for the most photometrically variable
stars near this limiting magnitude - as suggested by Figure C3b. The
limiting magnitude to which we can detect periods also might be
affected by whether the star is in a binary or not, since the light
from the other star in the system will act as an additional noise
source.  Since the single star bright end of our mass range in the
Pleiades (0.32 Msun) has a V magnitude only $\sim$2.5 mag brighter
than the limiting magnitude for detecting periods, it is plausible
that we might fail to detect periods in the secondary components of
photometric binaries if the secondary is too faint.

In order to assess this a little more quantitatively, we determined
system $V$ and \vmk\ magnitudes for hypothetical low mass Pleiades
binaries, using an empirical $V$ vs.\ \vmk\ single-star locus analogous
to that shown in Figure A3.   The primary in all cases had $V$ = 18.0
and \vmk\ = 5.4.   Secondaries ranged down to $V$ = 20.25, the faint
limit for our single star locus.  For an equal brightness system, the
combined light of the binary is shifted 0.75 mag above the single star
locus, as expected. However, even for the faintest secondary, the
predicted shift above the single star locus is $\sim$0.6 mag.   Thus,
it is entirely possible for many of the photometric binaries among the
$M <$ 0.32 \msun\ Pleiades stars to have secondaries that are simply
too faint to detect periods with K2.

\subsection{Conclusions Concerning Our K2 Binary Sample}

The $V$ vs. \vmk\ CMD for the Pleiades and Praesepe M dwarfs (Figure
2, lower panels) appeared to show a large population of photometric
binaries for which we only detected one period with K2.   In this
appendix, we have attempted to determine the cause for this apparent
anomaly.   Based on our analysis, three of the potential explanations
we have examined do not, in our opinion, significantly help to explain
this anomaly: (a) non-member contamination; (b) incomplete or
inaccurate identification of multiple periods in our K2 sample; and
(c) failure to detect a hypothetical population of very slowly
rotating companion stars due to their postulated low amplitudes of
photometric variability.  Instead, we believe our analysis has
demonstrated that there are two primary/likely explanations for the
apparent excess of stars displaced well above the  single star locus
in Figure 2 but having only one K2 period.  First, we believe that
errors in the \vmk\ colors have inflated the number of apparent
photometric binaries among the Pleiades dM stars, and that with better
photometry these stars often will fall within the single-star locus. 
Second, particularly for the lowest mass stars, if there is a range of
secondary to primary star mass ratios, many of the binary systems will
be displaced well above the single-star locus but have secondaries
that are too faint for K2 to detect their periods.  Based on this
analysis,  the apparent excess of photometric binaries that are K2
singles is unlikely to affect any of our conclusions with respect to
the rotation rates of single versus binary young, low mass stars.

\noindent
\end{document}